\documentclass[pdftex,12pt,fullpage,oneside]{article}
\pdfoutput=1 
\usepackage[top=1in,bottom=1in,left=1in,right=1in]{geometry}
\usepackage{setspace}
\usepackage{array,amsmath,psfrag,amssymb,tabularx,booktabs}
\usepackage[pdftex, hyperfootnotes=false, colorlinks=true, citecolor=dukeblue, linkcolor=dukeblue, urlcolor=dukeblue, breaklinks=true]{hyperref}
\hypersetup{%
    pdftitle={Irregular Leadership Changes in 2014},
    pdfauthor={Andreas Beger, Cassy L. Dorff, Michael D. Ward},
    bookmarksnumbered=true,
    citebordercolor={1 0 0},
    }
\usepackage{datetime}
\newdateformat{mydate}{\THEDAY~\monthname[\THEMONTH] \THEYEAR}

\usepackage{natbib}
\bibpunct{(}{)}{,}{a}{}{,}

\usepackage[T1]{fontenc}
\usepackage{fourier}		

\usepackage{graphicx}
\usepackage{color}
\definecolor{dukeblue}{rgb}{0, 0, 0.61}

\graphicspath{{/Users/cassydorff/Dropbox/RA/PITF/irregular-regime-change/graphics/}}
\graphicspath{{../graphics/}} 

\usepackage{dcolumn}
\usepackage{booktabs}
\usepackage{rotating} 

\newcolumntype{d}{D{.}{.}{-1}}
\usepackage{appendix} 

\usepackage[format=plain,justification=justified,labelfont=bf, textfont=it]{caption}
\usepackage[capposition=top]{floatrow}  
\usepackage{subcaption}   
\usepackage{multirow}
\usepackage{longtable}

\title{Irregular Leadership Changes in 2014: \\ Forecasts using ensemble, split-population duration models}
\author{Andreas Beger, Cassy L. Dorff, and Michael D. Ward \\
Duke University\footnote{This research was sponsored by the Political Instability Task Force (PITF). The PITF is funded by the Central Intelligence Agency. The views expressed in this report are the authors' alone and do not represent the views of the US Government. \newline \indent Version 1.3b, based on Archigos v3, compiled \today. Originally delivered on 8 May 2014.}}
\date{\mydate{\today}}
\usepackage{fancyhdr}
\begin{document}

\maketitle

\singlespacing
\begin{abstract}
\noindent 
We forecast Irregular Leadership Changes (ILC)--unexpected leadership changes in contravention of a state's established laws and conventions--for mid-2014 using predictions generated from an innovative ensemble model that is composed of several split-population duration regression models. This approach uses distinct thematic models, combining them into one aggregate forecast
developed on the basis of their predictive accuracy and uniqueness.  The data are based on 45 ILCs that occurred from March 2001 through March 2014, with monthly observations for up to 168 countries worldwide. The ensemble model provides forecasts for the period from April to September 2014. Notably,  the countries with the highest probability of irregular leadership change in the middle six months of 2014 include the Ukraine, Bosnia \& Herzegovina, Yemen, Egypt, and Thailand. The leadership in these countries have exhibited fragility during this forecast window.  \\

\noindent Forecast Map for April - September, 2014:
\begin{figure}[!h] 
   \centering
   \includegraphics[width=3.in]{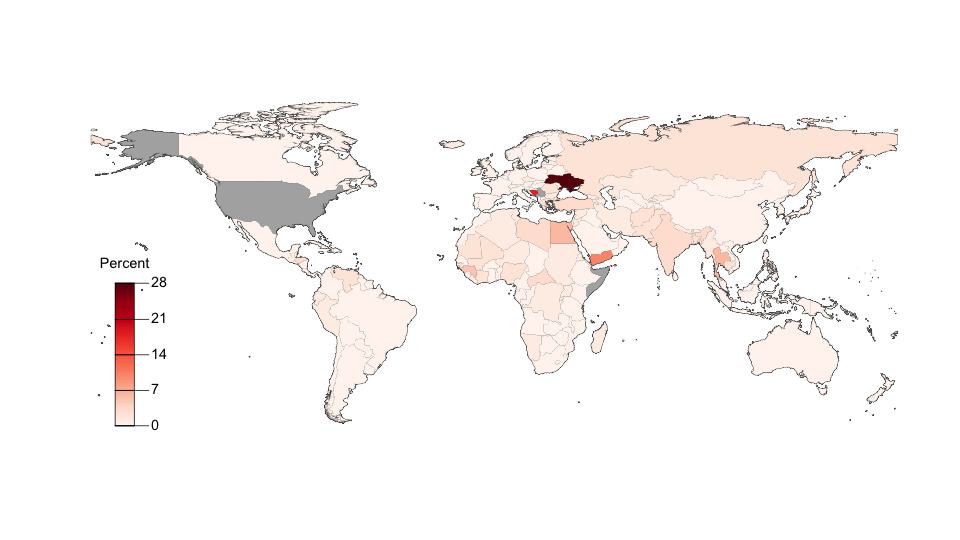} 
\end{figure}

\end{abstract} 

\clearpage
\tableofcontents
\clearpage

%
%
\section{Introduction}

In late February 2014, pro-Russian President Viktor Yanukovych of Ukraine fled the capital after mass protests erupted into violence, and parliament appointed an interim President to rule until May elections. The mass protests had broken out in November over Yanukovych's abandonment of an agreement for closer trade ties with the EU. A month earlier, in the Central African Republic, Muslim President Michel Djotodia was forced out of office in January 2014 in the face of escalating violence between the Muslim S\'{e}l\'{e}ka regime and the largely Christian anti-balaka coalition. The level of violence verged on genocide. Djotodia and the S\'{e}l\'{e}ka coalition had won power in March 2013 through a successful rebellion against the preceding government. 
In July 2013, the military in Egypt staged a coup and removed democratically-elected Mohammed Morsi from the Presidency, following waves of mass protest agains Muslim Brotherhood rule. 

These three events---a successful mass protest campaign, a successful rebellion, and a coup d'\'{e}tat---are treated as different types of events by most of the political science literature.
Yet they all share the same outcome, namely the sudden removal of a sitting regime. We call this outcome {\em irregular leadership change} (ILC): the unexpected removal of a political leader through means that contravene a states's conventions and laws. Irregular leadership changes are relevant for understanding the ebb and flow of political violence throughout the world. We shift the focus of inquiry from viewing ILCs as broken into distinct categories, and instead analyze leadership change as a general and politically relevant outcome. Thus, instead of addressing specific mechanisms that drive different types of ILC, i.e. narrow conspiracies, mass protests, or armed insurrections, we focus specifically on accurately modeling the common outcome of ILC. 

The goal of this project is to develop a model that can forecast irregular leadership changes, as catalogued in the Archigos dataset on political leaders. We use the concept of \textit{irregular} leadership change to capture those changes that occur outside of explicit rules or established conventions, e.g. coups, revolts, or assassinations \citep{goemans2009introducing,archigos2009codebook}.  We focus on modeling both the duration of a leader's time in office and the hazard that a leader's tenure comes to an end.  With a few exceptions such an approach has not been applied to these kinds of irregular political transitions and to our knowledge ours is the first attempt to use them for forecasting.

Examples of previous work using the Archigos data include \citet{gleditsch2010political} and \citet{gleditsch2004autocratic}. \citet{gleditsch2010political} show that irregular leadership transitions increase the risk of civil war by increasing the political opportunity for rebellion through reduced state repressive capacity and weaked leadership. \citet{gleditsch2004autocratic} model irregular leadership transitions in the context of political democratization. While these examples are tied to political conflict, other studies have focused more directly on political leaders themselves. \citet{besley2011democracies} use Archigos to examine whether democracies choose different political leaders than autocracies, and find that the former do indeed choose more educated leaders, which in turn seems to be associated with better economic performance \citep{besley2011educated}.

Our modeling framework allows us to draw on multiple arguments about different mechanisms that may lead to ILC, but ultimately our measure of success is how well we can predict. As part of this effort, we have produced out-of-sample global forecasts for the risk of ILC during the period from April to September 2014, which at the time the models were originally created in early May 2014 represents a 5 week lag. 

While irregular leadership changes appear, on the surface, to be related to coups, empirically there is incomplete overlap between empirical instances of irregular leadership change and government coups (see Section \ref{app:coupsILC}). Many coups are not classified as irregular leadership changes, and, conversely, many irregular leadership changes occur as the result of popular revolts or assassinations, not coups. Like coups, irregular leadership changes appear to be a heterogenous group of outcomes with different causal mechanisms. An implication is that the prior research on coups, including the arguments about causal mechanisms and specific empirical findings, are informative for modeling irregular leadership changes---but only to an extent. 

We  explore  the concept of leadership change by incorporating the insights from the literatures on several overlapping types of leadership changes: coups, revolts, assassinations, as well as changes that result from the direct threat of force by internal entities such as the military. In doing so, we examine a wide range of conditions under which leadership changes occur including the role of protests, ethnic fractionalization, and internal government conflict. 

To model ILCs we use split-population duration regression. This allows us to separately model the general risk of ILC in a country at a given time and the timing of an event for those countries that exhibit a high risk. We can thus separate slow-moving structural risk characteristics, like regime types and wealth, from more sensitive indicators of the particular timing of an ILC, like increasing levels of protest or signs of intra-government conflict.

Furthermore, given the broad range of potential explanations for different subsets of irregular leadership changes, we use an ensemble model consisting of several different split-duration models as input. Rather than collapsing various arguments into one kitchen-sink model with an array of unrelated variables, we develop several thematic models and aggregate them into an ensemble using Bayesian model averaging. This approach allows us to preserve interpretability of the input models while aggregating them into  superior forecasts that outperforms any of the individual thematic models. 

In the next section, we begin by describing what exactly constitutes an irregular leadership change; in doing so, we explore the specific differences between coups and the more general set of irregular leadership changes. Then, we review the relevant literatures and offer insights from each. Next we explain the research design, data, and split duration modeling approach. Last, we present our models and provide a discussion of each. In this section, we also provide forecasts for irregular leadership change in 2014. 

%
%
\section{What are irregular leadership changes?}

Irregular leadership changes are transitions between political leaders that occur outside the established rules or conventions of a state; this includes both the entry and exit of leaders. The Archigos dataset of political leaders \citep{goemans2009introducing} identifies, for each leader, whether he or she gained power through irregular means, and whether he or she lost power through irregular means. Irregular leadership changes are simply the composite of those two conditions, i.e. they occur when a leader gains power through irregular means, when a leader loses power through irregular means, or when a leader loses power through irregular means as a result of the succeeding leader's irregular gain of power, e.g. through a coup. 

The coding is specific to the circumstances by which political leaders gain and lose office, but the regularity of this gain or loss is relative to the established conventions, laws, and rules of a state governing succession between political leaders. Thus, in a democracy, a leader who leaves power after loosing an election does so in a regular manner in accordance with the state's conventions, whereas a leader who unwillingly resigns in the face of mass protests, without loss of an election and before a term limit is reached, exits an irregular manner. The definition of \emph{irregular} leadership change thus allows for variation in convention across states and is particular to the laws or norms of each state. A son who is given power by his father would be a regular leadership change in a monarchy, but could just as well be an irregular leadership change in a democracy in which transition is supposed to occur after elections. Many irregular leadership changes occur through coups or revolutions, but, unlike the latter two concepts, they are not defined by the means of leadership change. Rather, they are defined by the nature of transition in relation to the state's conventions and laws for transition. As a result, irregular leadership changes (ILC) do not quite fit into any one category of events that have been studied in political science, notably coups, rebellions, and revolutions.

\begin{figure}[h!]
\caption{Archigos irregular leadership exits, March 2001 through March 2014. Note that some countries had multiple irregular turnovers.}
\centering
\includegraphics[width=6.5in]{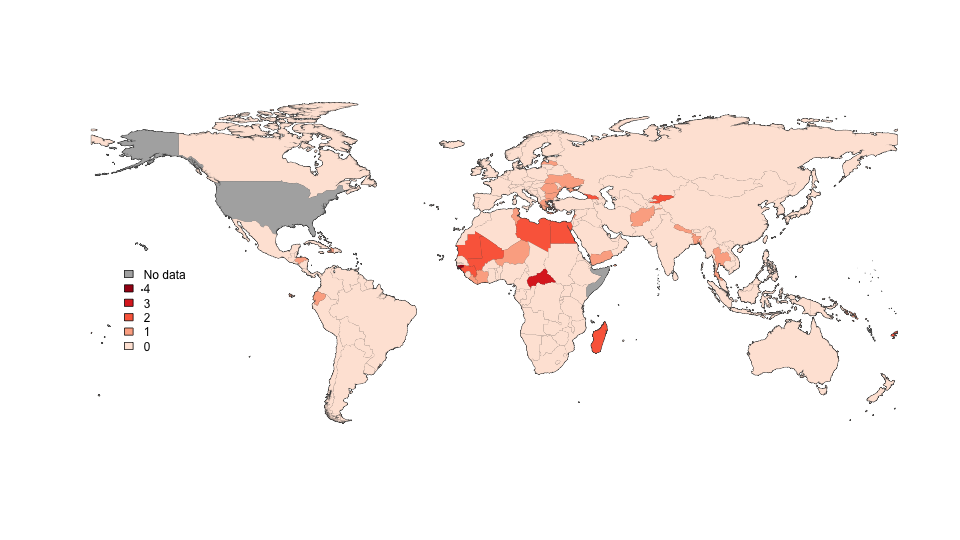}
\label{fig:irc}
\end{figure}

Since they are a relatively unusual outcome to study in comparison with existing research in international relations, we start with a broad overview of where and how irregular leadership changes occur. Figure \ref{fig:irc} maps all 45 irregular leadership changes that occurred between March 2001 and March 2014, the time period of our study.\footnote{Table \ref{tab:list} in the Appendix lists all ILCs.} They mostly occur in Africa and south and central Asia, with some outlier cases in Europe and Latin America. This pattern mostly matches that for other forms of political conflict, e.g. civil wars.

There are several different ways a leadership change can be coded as irregular. They generally fall into three broad categories: assassinations; revolts where the government is overthrown in the wake of mass protests or armed revolution; and coups whereby the government is overthrown by other parts of the state, usually the military. The majority of irregular changes are carried out by domestic forces without support from foreign actors, although this is not always the case. As an illustration, we briefly review three recent ILCs.

First, take the case of General Amadou Toumani Tour\'e, the former Malian President from 2002 to 2012. In March of 2012, the Malian military, displeased with the government's response to the Tuareg rebellion, took over the presidential palace in the capital city of Bamako. In doing so, the leader of the coup, Amadou Sanogo, and his military collaborators successfully forced the government of Amadou Toumani Tour\'e into hiding. This case qualifies as an irregular leadership change, with an irregular exit for Tour\'e and an irregular entry for of Sanogo at the same time. Both occurred as a result of internal government conflict and military pressure, and thus outside of regulated elections. This kind of classic military coup represents 35\% of the irregular leadership changes. 

Looking at the overlap with coups in more detail, there are 16 coups during the period from March 2001 to December 2013, compared with the 40 irregular leader exits.\footnote{Using data from \citet{powell2011global}.} Of those 16 coups, 15 are also coded as irregular leadership changes in our data. The exception is 2005 Togo where President Eyad\'{e}ma died on 6 February 2005, and the military installed one of his sons, Faure Gnassingb\'{e} before the President of the National Assembly, next in line for succession, could return to the country. Although controversial, Gnassingb\'{e} later won elections, and the case is not coded as an irregular entry in the Archigos data, and thus not as an irregular leadership change.\footnote{This time period is covered by the original Archigos coding, not our own update of the data for 2012 through the present.} Coups are thus a subset of irregular leadership changes.

The second example is in Romania, where Emil Boc became Prime Minister after being able to build a majority coalition in the wake of the 2008 legislative elections. His government lost a vote of no confidence in October 2009, but, following a narrow victory of President B\u{a}sescu, he was reinstated. The President himself had earlier been suspended from office in the course of a failed impeachment attempt on matters related to government corruption. In addition to corruption, the government became increasingly unpopular as a result of a series of austerity measures that were introduced to obtain an IMF loan in 2009. The loan became necessary in the aftermath of the 2008 financial crisis and at a time when GDP was contracting by more than 6 percent. Changes to health care laws and dismissal of a critical health minister finally led to open protests agains the government in early 2012. These events initially led to Boc's resignation and later cumulated in a second attempt to impeach the President. Boc's resignation is coded as an irregular exit because it occurred under protest pressure, and thus is coded as an irregular leadership change. Irregular leadership changes as a result of mass protests represent a further 38\% of irregular exit cases.

Another example is Laurent Gbagbo's irregular exit in 2011. Gbagbo was president of C\^ote d'lvoire from 2000 on, after winning an election and after street protests forced his reluctant predecessor to recognize the results and leave office. Originally elected for a five year term, a civil war led to repeated postponement of new presidential elections. When elections finally took place in 2010, Gbagbo lost but refused to leave office. Fighting broke out with opposition forces, and in April of 2011 he was deposed and ultimately arrested by these so-called rebels, with some participation by the French military forces. Shortly thereafter, he was taken into custody by the International Criminal Court under allegations of war crimes committed during this conflict period. These types of revolts, along with coups conducted by non-military actors, make up most of the remaining ILC cases. 
 
As these cases show, there is a wide variety of events that can lead to an ILC. The composition of such a heterogeneous set can in large part be explained by the source data, Archigos.  Archigos is focused on political leaders and the circumstances of their entry, duration, and exit from power. Using it to study abnormal government changes is somewhat incidental. Generally, studies of political events that result in leadership change are instead grouped by the mechanism that leads to the change, distinguishing coups undertaken from within government against a leader, from mass protests undertaken by the larger population outside of the government. This makes sense insofar as these phenomena reflect different underlying causal processes. ILC is instead focused on the common outcome, which is unexpected and abnormal leadership change. In regard to our study, the empirical heterogeneity suggests that irregular leadership changes are a multi-causal phenomenon. The factors that cause classic military coups, like those in Mali in 2012, are likely to be different from the widespread dissatisfaction and other factors that lead to popular revolutions, like in the Ukraine in February 2014. Thus, for the purpose of informing a modeling effort, both the literature on coups and revolutions are relevant and we briefly review them next.

\subsection{Coups}

Early interest in coups began in political science during the 1960s \citep{huntington1968political, jackman1978predictability, johnson1984explaining}. Inspired by a wide range of coups that took place in Africa, in places like the Congo (1963), Algeria (1965), and Uganda (1971), these works focus on the structural determinants of coups. \cite{jackman1978predictability} finds that social dynamics, namely mass citizen political turnout and the presence of a dominant ethnic group, are destabilizing. In contrast, \cite{johnson1984explaining} find that this does not apply well to military coups and that instead economic factors and the degree of politicization of the military best predict coup occurrence. They note that cohesive militaries with traditionally strong roles in domestic politics are more prone to stage coups in response to perceived political or economic crises. \cite{mcgowan1984} make an important contribution by including attempted coups in their analysis and conclude that the driving force behind coups in Africa during the 1956-1984 time period is best explained by the instability brought on by failed attempts at industrialization. 

Summarizing this earlier literature, \citet{goemans2011elections} note three distinct classes of arguments for why coups occur: political instability resulting from rapid economic modernization \citep{deutsch1961social,huntington1968political}; political illegitimacy following lackluster economic performance and development \citep{mcgowan2003african}; and conditions that increase the likelihood of military intervention in politics \citep{johnson1984explaining, jenkins1990explaining}. We do not see these arguments as necessarily disjoint: while one set informs us about the conditions under which a coup might occur, e.g. as a result of certain structural conditions like the political system, factionalism, or a politicized military; the other set provides traction on {\em when} a coup may occur if the structural conditions are ripe.

More recent work has focused on coup-proofing and leader decision making in an environment where coups are a threat. \citet{goemans2008way} focuses on leadership change from the perspective of the leader. Specifically, he suggests that leaders are fully aware of the dire consequences that can await them after poor performance in office. For example,  \citet{goemans2008way} finds that leaders who are defeated in external wars are more likely to face an irregular removal from office. Since leaders know such consequences await, they strategically behave in a manner that best secures their survival. Thus, leadership performance is an important component in our understanding of the baseline risk to leadership removal. 

\cite{de2005logic} and \cite{svolik2012} also approach irregular regime transitions from the leader's decision making perspective. Svolik's work explores the ever-present tension between the dictator and his ruling elites. He characterizes this dilemma in terms of violent threats and a shifting balance of power between key actors: as a dictator consolidates power he might choose to eliminate key elites that potentially pose a threat to him. The decisions that the dictator makes are imperfectly observed by his ruling coalition. If the dictator does begin to shift power, and in-doing so decides to eliminate key elite players, this increases the level of threat perceived by the ruling coalition. As this threat increases, so does the likelihood that the elites would take a risky and costly chance at a coup. 

Congruently, \cite{de2005logic}  argue that a leader's survival in office is centered on their the role of supporting coalitions within the government. Using similar logic to \cite{svolik2012}, this approach argues that political leaders establish power and influence by having the support of the winning coalitions, which can be thought of as either a select group or a portion of total citizenry. This winning coalition chooses the leader, and also provides the leader with the support he needs to rule. The relationship between the leader and his supporting coalition is a continuous two-way street: in order to maintain power and survive in office as long as possible, the leader allocates special goods to his supporters (and thus winning coalitions are known to be smaller in autocracies  and larger in democracies). If discontent arises within these supporters the likelihood of a coup can be described as a function of the size of the winning coalition.  

Despite that, to some degree, studies on different types of leadership change have similar theoretical claims, little attempt has been made to integrate this body of knowledge and leverage it for prediction analysis. Furthermore, the two part logic driving this literature has gone empirically unexplored. We will explain and demonstrate how the phenomenon of irregular leadership change, including coup events, can (and should) be modeled using split-duration models.  

This two part logic is summarized as follows. The first part outlines a form of structural risk produced from within a state's institutions, which is captured by the history of political violence the state experienced. The second part of the story focuses on the types of events that trigger instability and coups in countries. For example, \cite{belkin2003toward} explicitly distinguish the structural risk of coups from specific triggers that determine the timing of a coup in their causal story. They argue that a state's vulnerability to a coup can be captured by measures on civil society, the government's legitimacy, and past coup history. Such measures allow the authors to analyze the ``deep'' rooted structural attributes of the government, its citizenry, and the interaction of the two. Short term crises, however, hastens the occurrence of a coup. Similarly, \cite{galetovic2000} find that economic recessions increase the likelihood of a coup. 

Continuing with this dual theoretical story, in research demonstrating the utility of coop-proofing, \cite{powell2012determinants} argues that the role of the military is key for understanding coups, in that its size relative to the population, general soldier satisfaction, and organizational cohesiveness help predict whether the military will decide to depose the civilian government. This argument fits well with the empirical cases we've already mentioned as examples of military involvement as a key influence in leadership change. Powell, like others \citep{koga2010, galetovic2000}, also notes that if the status quo is threatened through shocks like economic crises, even the most satisfied militaries may view coups as favorable. 

Much of the intuition in the coup literature suggests a two part modeling approach whereby one aspect of the model can specify the baseline risk of a coup and the other captures triggering events. More recent thinking regarding coups thus matches well with a split-duration model in which conditions that determine the risk of a coup occurring at all are separated from conditions that determine the timing of a coup, i.e. triggers. At current, a majority of extant modeling relies on binary response models that do not capture this theoretical distinction directly.

\subsection{Revolutions}

The wave of revolutions that brought down communism in Eastern Europe in the early 90's and the wave of revolutions during the Arab Spring in 2011 have each generated a large interest in explaining how these revolutions could have so unexpectedly affected monumental change in once ``stable" regimes, and whether it is possible to predict them \citep[e.g.][]{kuran1995inevitability}. Explanations for the suddenness and apparent unpredictability of such revolutions have focused on tipping points that lead to cascades of protest, which effectively overwhelm the established state institutions \citep{kuran1991now}. In this vein of research, the causal mechanism story presents as follows: there exists a segment of the population in an autocratic state that would participate in a revolution to overthrow the regime \textit{if} the revolution is going to be successful, but for understandable reasons, the population does not want to support a failing revolt. The initial mass of protesters thus has to reach a tipping point that will successively persuade more and more people that a revolution is coming and hence to join, until leadership change is inevitable. This logic is at the root of the research on civil resistance \citep{chenoweth2011}, which suggests not only that numbers matter but that defections by those close to central power--such as military members-- is key for a revolution to succeed. 

As the tipping point argument makes clear, some initial protest activity is necessary to spark a revolution. At the same time, many regimes have norms allowing for low levels of peaceful protest, and protest activity in a country itself may therefore not be indicative of an impending leadership change. Studies on the timing of revolutions build off of previous studies that identify those structural factors that create the initial conditions for a revolution \citep{goldstone2001toward}. In our modeling framework, which we discuss more below, these slower moving, structural conditions correspond to factors that determine whether states are in the ``risk set'' for revolutions.

The military is another key player within the dynamics of a revolution. How the military responds to potential threats and mass protests is important to the way that these conflicts evolve. In Tunisia, the head of the military received orders from the president to shoot protesters. Instead, he placed his forces between civilian protesters and paramilitaries supporting the president, effectively removing the president from power \cite[31]{barany2011role}. In Bahrain, the military and security services, with help from neighboring states, effectively suppressed mass protests against the governing monarchy. 
In many autocratic countries facing mass protests, the military eventually gets the order to shoot \citep{barany2011role}. Some do, others do not. In Syria, it was even more complicated as large parts of the military proved willing to shoot at demonstrators, but others defected rather than fire on civilians.  In the end, the situation turned into a violent intrastate conflict. 

\citet{barany2011role}, examines the role of militaries in countries experiencing unrest as part of the Arab Spring, and offers three factors that play a role in the military's decision: professionalization, the role of the military in the current regime vis-a-vis  other security services, and the potential impact of a successful revolt on the military's own interests. More professional militaries, with exposure to foreign training and relatively low political involvement in state affairs, should be more hesitant to suppress peaceful mass protests than militaries whose institutional basis is more closely associated with a ruling government and focused on internal security. In terms of the military's interests, a higher level of professionalization should also be indicative of institutional independence from the government. Looking further at the military's interests, \citet{barany2011role} notes that the military is not willing to repress protesters in countries where the military is sidelined by other internal security forces. Finally, armed forces closely tied to a government may be more willing to support the government by all means, including violent suppression, if a successful revolt would be detrimental to the military's role, funding, and other interests under the successor regime. In countries like Bahrain and Syria, where the military supports an ethnic minority regime, and officers tend to come from those minority groups themselves, a successful coup or revolution would likely lead to an environment unfavorable to the military. Thus, in addition to those factors that may encourage citizens to partake in mass protests, like poor governance, the military's behavior is a key determinant of a revolution's success. 

\section{Synthesis and approach to modeling}

Coups carried out by one part of the state against another are without a doubt very different from revolutions affected by mass protests. Both literatures, however, have common themes that we used to guide our modeling approach: unity, or lack thereof, in government and dissident actors, the role of the military, and factors that distinguish structural conditions from immediate triggers. 

Much of the literature on coups and revolutions is constrained by data availability to structural factors that affect the general susceptibility of a country to an event. Structural factors, in this context, are those that are static or slow-moving over time, like GDP, or government budgets. In a more conceptual realm, this includes the sorts of factors that might lead to widespread dissatisfaction with a political system and eventual revolution, or those indicating a politically-involved military. We draw on a fairly broad range of indicators for such structural conditions through the ICEWS data, described in more detail below. Aside from practical constants, there is a clear theoretical position for structural factors indicated in part by the fact that no sensible person would expect a military coup in Sweden, but may not be surprised at a military coup in Egypt, based on the fact that Sweden is an established, wealthy democracy, while Egypt is an autocracy with a politically active military and many economic sources for discontent. 

But slow-moving forces do not help explain the specific timing of irregular leader exits. The coup literature specifically mentions this bifurcation between structural risk and immediate event triggers \citep{belkin2003toward}. One of the fundamental questions underpinning any discussion of revolutions, like those of the Arab Spring, is in understanding their timing and sudden onset, despite years of relative peace in stagnant systems. To some extent there are explicit claims about what might indicate an immediate event, e.g. that loosing a war may spur leadership change, like the loss of the Falkland War in 1982 did to the military junta in Argentina. Another approach is to look for early, dynamic indicators in addition to structural ones. For example, the kind of open, public conflict between the military and the government could be an early indication of a military coup. Preventing unity among opposing groups is a fundamental strategy exploited by political leaders, and in fact one of the main measures of coup-proofing autocratic regimes \citep{quinlivan1999coup, pilster2011coup}. Exploring the public discourse between dissident actors and government actors can serve as an indicator of fragmentation between these different sectors in a society. Part of our modeling effort builds on the creation of such indicators from event and other data.

Beyond the need for time-varying indicators to pin down the timing of ILC in countries that are likely to be at risk, a second major point to note is the large number of potential mechanisms that may lead to an ILC. Not only is ILC conceptually heterogeneous, encompassing coups, armed rebellions, and mass protests, but within each of these facets of ILC there is a range of viable arguments that can serve as the basis of a model. As explained further below, our modeling approach reflects this complexity. After describing our data and methodological approach in the following sections, we then expand on the details for each thematic model and present our results.

Finally, a third point to note is that while the literature on leadership change, along with our own previous empirical modeling efforts, suggest a wide range of factors and related variables that may be useful in a model of irregular leadership change, ultimately, a set of ideas that cannot predict well is not very useful. \citet{ward2010perils} show that the statistical significance of covariates in a model is not a good guide to the predictive performance of that model, and many well-accepted conflict models do in fact have rather bad fit to the data. Conversely, covariates that lack statistical significance can make large contributions to model fit. Two thoughts emerge from this: first, that a set of ideas is good to the extent that it generates predictions that match the observed world, and two, that statistical significance is not a good guide for creating a well-fitting model. Our modeling efforts as a result reflect a mixture of theoretical and empirical guidance. Theoretical guidance to suggest sets of covariates that are relevant, which is especially helpful in crafting indicators from event data, and empirical evaluations of model fit to guide variable selection and model importance.

Our modeling is thus informed by the need for prediction, which in turn creates the need for time-varying data that is measured below the country-month level, and most importantly, by an effort to accommodate the heterogeneity of possible mechanisms and related models that may be tied to ILC. To this end, instead of creating a kitchen-sink model we introduce 7 thematic models:
\begin{enumerate} \itemsep0pt \parskip0pt \parsep0pt
\item Leader characteristics: this thematic model captures the leader's individual characteristics and regime conditions, such as length of time the leader has been in power and intra-governmental cooperation. 
\item Public discontent: this model focuses on public discontent within government and between government and dissidents as an early indicator for ILC in societies with strong media use. 
\item Global instability: as we mentioned in the previous section, studies focusing on leadership change have focused on structural variables. To best capture this contribution, we replicate a model based on the \citet{goldstone2010global} model.
\item Protest: it is evident that mass protest are often an effective means to destabilize the regime. Our protest model captures low level indicators of conflict--such as strikes, barricades, and intra-ethnic conflictual events. 
\item Contagion: this theme represents the idea that conflict, and especially protest events, work through a contagion-like process. To reflect this, our contagion model includes two spatial weights of opposition resistance and state repression in neighboring countries. 
\item Internal conflict: conflict by itself can be a mechanism to affect ILC through successful armed rebellions. In addition, other types of low-intensity conflict may have a shielding effect by allowing a leader to shift focus to security issues, rather than legitimacy.
\item Financial: an often neglected aspect of regime stability are it's underlying finances. This theme focuses on inflation and international reserves as triggers and buffers for ILC, respectively.
\end{enumerate}
We fully describe the themes and what they entail  after first introducing our research design and some technical details of our modeling strategy.

%
%
\section{Research design}

Our dependent variable is a binary indicator of whether an ILC occurred in a country in a given month. To model the outcome of interest we use a set of split-population duration regression models, each motivated by a particular substantive theme, which are combined into one ensemble model that provides the final forecasts. These models are built on the basis of ILCs between March 2001 and March 2014. These date limits are driven by availability of data from the ICEWS project, although we partially use earlier data on political leaders, as discussed below, to ameliorate left-censoring. The basic unit of observation is the country-month and we observe a data point for up to 168 countries worldwide for each of the 157 months covered by our data. 

Since the overall process is more complex than conventional modeling with panel data, from building the data to generating the final forecasts, we summarize the steps below.	
\begin{enumerate} \itemsep0pt \parskip0pt \parsep0pt
\item Build duration data, i.e. time to failure counters and risk/cured classification of spells using ILCs that occurred in the international state system from 1955 through March 2014.
\item Add covariates from PITF, ICEWS, and other sources to the data from March 2001 on, and drop data before this date.
\item Estimate thematic split-population duration models using training data from March 2001 through December 2009. 
\item Calibrate an ensemble ILC models using predictions from the input models for January 2010 to April 2012. This provides the models weights that optimally combine the individual model predictions.
\item Test the ensemble ILC model predictions, i.e. weighted combination of input model predictions, on observed outcomes from May 2012 through March 2014 in order to estimate the model's out-of-sample predictive accuracy.
\item Generate $n-$month ahead forecasts using the ensemble predictions based on data from March 2014.
\end{enumerate}

The rationale underlying our choice to aggregate several split-pop\-ulation duration models into one ensemble model is the heterogenous nature of our outcome variable.  Not only does irregular leadership change have multiple facets, but each of these facets is plausibly caused through several distinct mechanisms. The mechanisms that led to successful protests in Egypt are likely to be different from the mechanisms that lead to successful protests in Ukraine. Rather than collapsing these different arguments and explanations in order to build one unified model, we build several input models, each capturing a different argument, and aggregate them into one ensemble model that draws on the strengths of each input model for a superior final forecast. 

\begin{figure}[htbp] 
   \caption{Data partitions}
   \centering
   \includegraphics[width=6.45in]{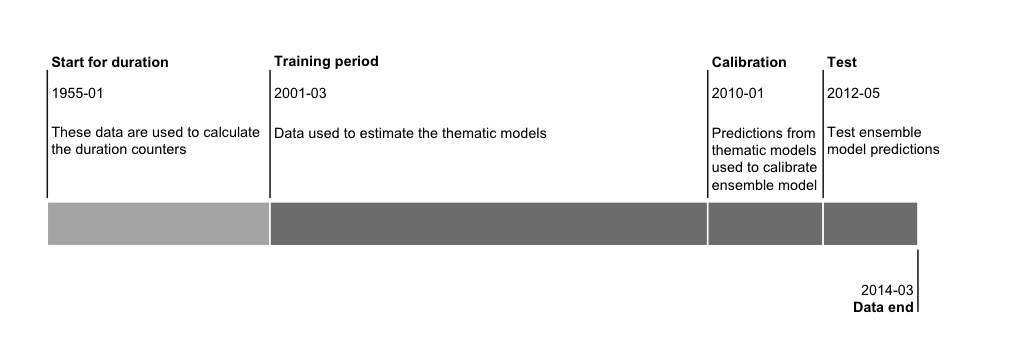} 
   \label{fig:design}
\end{figure}

Using this modeling strategy requires us to partition the data into several different sets, as shown in Table \ref{fig:design}. In the next few sections we introduce each partition set along with the technical concept, e.g. split-population duration regression and ensemble model averaging, that it supports. 

First, although our covariate data are only available from March 2001 on, we use data from Archigos reaching back to January 1955 to build variables that count the time between failures (irregular leadership changes). This is to ameliorate left-censoring problems, which occur when we do not accurately observe the time from a previous failure to one captured our data because the previous failure occurs before March 2001. Since a counter of the time between failures is a key component of duration modeling, this can lead to inaccurate model estimates. Using 1955 as a start date for building this counter provides us with $\sim$700 extra months of data in addition to the 157 captured by our main data to inform the time to failure counter, and thus solves the left-censoring issue for practical purposes. 

We should further note that all research that relies on panel data, i.e. countries observed over some time period, also suffers from the same left-censoring problem. Splines or time polynomials rely on a time to failure counter like we do, and the alternative of ignoring temporal dependence hardly seems like a solution.

\subsection{Covariates and event data}

Our data include approximately 200 potential covariates which fall into 3 broad categories. The first are structural variables like GDP per capita, the Amnesty Political Terror Scale, or regime types. These variables tend to be measured at the country-year level, and they mostly vary between countries rather than within any particular country. Thus they are more useful for distinguishing risk sets than predicting the timing of particular events. 

The structural variables include several economic and financial indicators like GDP, population, mortality, military expenditures, broadband subscribers, cell service subscribers, foreign direct investment, and CPI from the World Development Indicators \citep{world2013world}, the Polity regime variables \citep{marshall2002polity}, indicators for the number and power relationships of ethnic groups from the Ethnic Power Relations data \citep{cederman2009ethnic}, the Political Terror Scale \citep{wood2010political}, as well as secondary measures constructed from the Archigos data, like indicators for leaders who entered irregularly or through foreign imposition \citep{goemans2009introducing}.

The second group of variables, which we call behavioral, are constructed from the ICEWS event data, and record the number of certain types of events in a country over the course of a month, e.g. protests directed at the government. The \href{http://www.lockheedmartin.com/us/products/W-ICEWS/iData.html}{ICEWS event data}, like GDELT, \href{http://www.acleddata.com/}{ACLED}, \href{https://www.strausscenter.org/scad.html}{SCAD} and \href{http://www.pcr.uu.se/research/ucdp/datasets/ucdp_ged/}{UCDP GED}, are based on (machine) coded media reports, which are parsed for actors, locations, and actions to create distilled event records.  See the appendix for a more detailed discussion of the ICEWS event data, including plots of events over time and by country.

We include various aggregations of the ICEWS event data, particularly so-called quad variables that capture verbal and material conflict and cooperation within government, and between government and dissidents and vice versa. For example, verbal cooperation includes making positive public statements, appeals, or consultations, while verbal conflict captures reports of investigations, public demands, or threats. The complete mapping between these categories and specific types of events is detailed in the appendix, specifically Table \ref{tab:quads}. Other aggregations capture protests towards government, as well as incidents of rebellion and insurgency. These variables vary somewhat between countries, but mostly change over time within countries, making them useful for timing the onset of events. 

The third set of variables are spatial lags of the behavioral, event-based variables. A spatial lag in essence captures what is going on in the neighborhood of a country, e.g. what the average level of protests in Egypt's neighboring countries at the time of the uprisings was \citep{ward2008spatial}. There are different ways to define what constitutes a country's neighborhood, and we include weights constructed on the basis of the 4 nearest neighboring countries, the distance between country centroids, and finally, Gower distances \citep{gower1971general} of country's similarity on either political, economic, or event measures. The latter capture distance as the dissimilarity of countries' political regimes, for example, rather than spatial distance. Using these different weighting methods, spatial lags are then constructed for rebellion, insurgency, and other similar behavioral variables. The spatial lags tend to vary more over time than between countries, since they by definition average out some between-country variance. 

\subsection{Split-population duration regression}
The total number of observation in our data is approximately 26,000, and with 45 irregular leadership changes in the non-missing data. This equates to a positive rate of 17 hundredths of a percent. The second partition, from March 2001 through January 2010, forms our training data. These data are then are used to estimate the split-population duration regression models that serve as inputs to the final ensemble model. 

The thematic models are based on split-population duration regression \citep{svolik2008authoritarian}. Basic duration models were initially developed in a health context to examine the survival of medical patients and to estimate the risk of an event like a heart attack at a given time, given that an individual has survived up to that point in the first place.  We can similarly use them here to model the survival of stable regimes given some risk of irregular leadership change.

An important distinction that motivates split-population duration models is to realize that not all polities are at risk of failure. For all practical purposes, countries like Germany, Canada, or Japan are highly unlikely to experience ILC within the time period we are interested in here, whereas many countries in Africa and the Middle East do experience observed instances of ILC over the past decades. The important point from a modeling perspective is to conceptually separate countries, or in this case country-months, which are ``at risk'' of failure. Additionally, we must separate out those countries which are practically immune from ILC.  

\begin{figure}[htbp] 
   \caption{Country A is at risk; Country B is not at risk. Conflating these in
duration models provides too low a risk assessment, and overestimates the decline in risk over time.}
   \centering
   \begin{subfigure}[b]{0.55\textwidth}
   	\includegraphics[width=\textwidth]{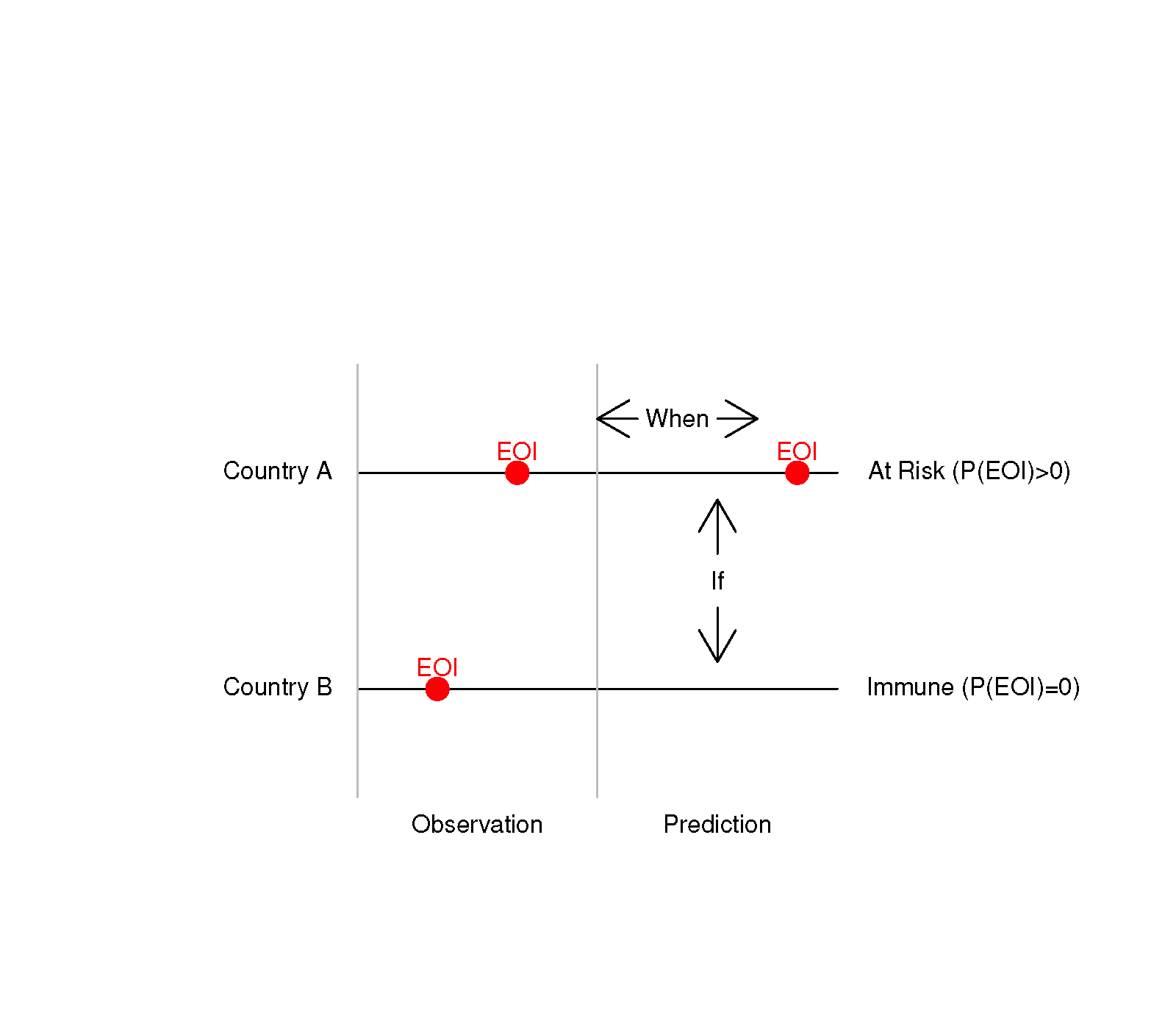}
	\caption{Splitting Population}
   \end{subfigure}
   \begin{subfigure}[b]{0.4\textwidth}
   	\includegraphics[width=\textwidth]{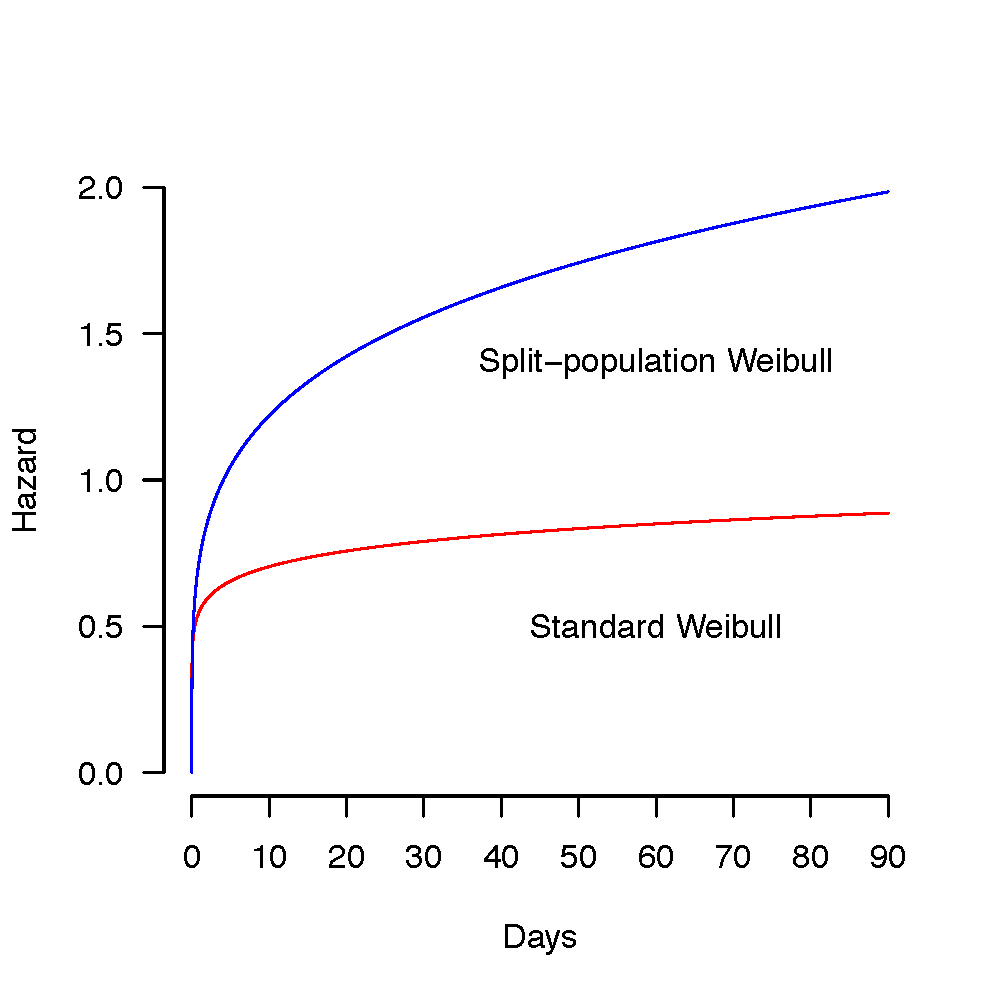}
	\caption{Baseline hazard}
   \end{subfigure}
   \label{fig:sp}
\end{figure}

Figure~\ref{fig:sp} illustrates the intuition behind this approach. As shown in the left panel in Figure~\ref{fig:sp} there are two types of polities. First are those that may have had an event, but essentially are immune from further events (Country B). These include countries that never had any events, but are not shown in this illustration. The second type of polity is at risk for future events (Country A).  The split-population approach models first the separation of locations into type A or B; denoted by the {\em if}~~ in the panel. The next part of the model determines the duration of time until the next event, denoted by {\em when}. The right panel illustrates the differences in base hazard rates under the assumption that all locations have the same risk profile (the Standard Weibull) compared to the baseline risk that assumes the population of locations consists of two types: those at risk and those immune from risk.

The basic likelihood of this kind of situation has been completely worked out, and may be thought of as a mixture of two likelihood functions.\footnote{
The likelihood is given as a product of the immunity and the risk: $
\mathcal{L}\{\theta|(t_{1}, \dots, t_{n})\}=\prod_{i=1}^{N} \left\{(1-\pi) f(t_i)\right\}^{\delta_i} \times  \left\{\pi + (1-\pi)S(t_i)\right\}^{1-\delta_i}
$
} This likelihood function reflects a mixture of two equations: a first step in which it classifies risk and immunity, and a second step that models expected duration to failure. One advantage of this modeling approach is that it allows covariates to have both a long-term and short-term impact, depending on which equation they enter in the model. Variables that enter the immunity equation have a very long-term impact because they change the probability of being at risk at all. Variables in the second, duration equation can be thought of as having a short-term impact that modifies the expected duration until the next failure.

Data for split-duration (and duration) modeling, in a country-month framework, conceptually consists of ``spells.'' Spells are not unique to countries, and instead consist of the time in a country from a previous irregular exit (``failure''), or the date the country came into existence, until the next failure or until the censoring time is reached. The censoring time is the last month of data we observe, December 2013 in our case, and indicates that data for a spell has ended, but not because of failure (``right-censoring''). 

Another important thing to note in the context of country-month data is that we model country-months as being in the risk set, i.e. a country at a particular point in time. As the inputs change over time for a country, so can the estimate of whether it is in the risk pool at any given point in time. So, while it is helpful on a conceptual level to speak of countries as being at risk or immune, in a more detailed, technical sense we should speak of a country at a point in time. Canada may not be at risk in 2014, but it may be at risk in 2015 if some unanticipated disaster leads to conditions that we associated with country-months in the risk pool. 

To use the country-month data with a split-duration model, we need, among other variables, a counter for each spell that indicates the number of months since the start time or previous failure time. This introduces an issue when we do not observe the previous failure time, i.e. ``left-censoring''. If we were to start all counters in 2001, the US in 2004 would have the same counter value as Serbia, which had an irregular exit in October 2000 (Milosevic). To mitigate this problem, we use 1955 as the start date for building the duration-related variables. We include the information from Archigos for this time period. Therefore in our data starting in 2001, the counters will have started either at the date of the last previous irregular exit for countries that had an irregular exit between 1955 and 2001, or at 1955 for those that have not.\footnote{Note that other types of models, e.g. probit, suffer from the same left-censoring problem if they attempt to control for temporal dependence by using variables like indicators for previous irregular leadership change or cubic splines. }

The other duration related variables we need are an indicator for right-censoring, when spells end because we have reached 2013, and a binary indicator for spells that are considered to be at risk of failure. According to convention \citep{svolik2008authoritarian}, we retroactively code a spell as being at risk if it ended in failure. Spells that are right-censored are coded as being not at risk, as would be spells that end when a country ceases to exist. Treating right-censored spells as not at risk can be problematic since they may later, after we observe more data, end in failure, meaning we miscoded them in the first place. We are not aware of an alternative to this coding scheme however, and given the low number of spells that end in irregular leader exit, it is probably safe to assume that right-censored spells are mostly not at risk.

\subsection{Interpretation of split duration estimates}

Once estimated, there are several different quantities we can calculate from a split-population duration model. The one we focus on is the \textbf{conditional hazard rate}, which estimates the probability that a country will experience failure during a given time period, considering that it has already survived without failure up until then, and considering that some countries will never experience failure. 

To fully understand this we can break the conditional hazard rate for a country at a time $t$ down into it's two components: the unconditional hazard rate, and the probability that it is at risk of failure (i.e. not cured) at all. To provide a specific example, we will focus on Mali, which had a coup in March 2012, and assume that we are modeling the outcome at a monthly level. But the same can be more generally said about arbitrary subjects or countries over arbitrary time periods, with different types of failure events.

The \textbf{unconditional hazard rate} is the probability that a country will experience a failure in a given month, given that it has survived without failure up until that month. We can estimate it based on historical data on previous coups in other countries and previous coups in Mali itself, and the time it took until these events happened in each case. Thus for Mali, the unconditional hazard rate for February 2014 would give us the probability of an irregular leadership change in Mali, given that it will have been 23 months at that point since the previous coup in March 2012. We could just as well calculate the hazard rate for any other month, like January 2014, or April 2014.

There are three things to note about the hazard rate. First, it is specific to a given country and a given time since the last event. The hazard rate for irregular transfers in Mali is likely to be different from the hazard rate our model estimates for Chad. Similarly the hazard rate for  in Mali 6 months from the last irregular transfer will probably be different from the hazard rate at 12 months from the last one. Second, these changes in the hazard rate over time can follow specific shapes. They can be flat, meaning that the hazard of an event does not change over time. This is the case for example for the decay of radioactive elements. But the hazard rate for irregular transfers could also have a bump in the beginning, indicating that these are more likely a short time after a transfer has already happened, but less likely over time as regimes consolidate power. In any case, a variety of shapes are possible. 

The third thing to note is that this unconditional hazard rate assumes that all countries are subject to experiencing irregular transfers, including coups. That hardly seems sustainable, and for all practical purposes we would not expect countries like Sweden or Switzerland to experience coups at all during the time frames we are interested in. To that end a split-population duration model also tries to group countries which are at risk from experiencing a coup, like Mali, and those that are effectively ``cured'' of coups like Sweden or Switzerland. 

The \textbf{risk probability} is an estimate that a given country at a given time falls into either the group of countries at risk ($p=1$), or the cured group ($p=0$). We say that it is \textit{an estimate} because while countries like Mali, Sweden, or Switzerland might be easy to classify, other countries like Ukraine, Nigeria, or Venezuela are less clearcut and fall somewhere in between. We say that it is an estimate \textit{at a given time} because it also depends on how much time has passed since the last event. Our guess about whether Mali is cured of coups is surely different when it has been 24 months since the last coups than if it was 240 months since the last coup.

To calculate the conditional hazard, we combine our estimated hazard rate with the estimated probability that a country at a given time is in the ``at risk'' set of countries. In other words, \textbf{conditional hazard} $=$ \textbf{unconditional hazard} $\times$ \textbf{risk probability}. For Mali, we estimate the probability of an ILC in February 2014 to be the unconditional probability of a coup when it has been 23 months since the last event times the probability that Mali is in the ``at risk'' set of countries given it's characteristics and given that it has been 23 months since the last coup. In this way we can get a probability estimate for a coup that takes into account the changing hazard of coups over time, but which also corrects for the fact that some countries will never experience a coup.

The \textbf{coefficient estimates} can be interpreted similar to standard regression coefficients. The sign of coefficients in the risk equation indicate whether a change in a variable increases or decreases the probability that a country-month is in the risk set, and exponentiated coefficients indicate the factor change in risk probability associated with a 1-unit change in the associated variable. The duration part of the model is in accelerated failure time format (AFT), and for interpretion it is convenient to think of the dependent variable as being survival time, or, equivalently, time to failure. A negative coefficient shortens survival and thus hastens failure (higher probability of an event at time $t$), while a positive coefficient prolongs survival and thus delays failure (lower probability of an event at time $t$).

\subsection{Variable variance and risk estimation} 

On an intuitive level it is easy to think of risk as being particular to countries, but in a technical sense risk is specific to our unit of observation, country-months. One implication of this is that if the inputs to the risk equation are time-varying within countries, e.g. measures of the presence of various types of conflict, the risk probability for a single given country will also change over time. Conversely, although it is intuitive to think of the duration equation as predicting timing, it is entirely possible to build a well-fitting duration component using static, structural variables. 

To ensure consistency with the interpretation of a split-population duration model in the context of country-month data that we have presented above, i.e. that the risk equation distinguishes countries at risk of ILC from those immune to ILC, and that the duration equation models the timing, we restricted variables that enter the risk equation to \emph{static} variables that change primarily between countries, and less within countries over time. Examples include GDP per capita and Amnesty International's Political Terror Scale values. The duration equation includes the remaining \emph{dynamic} variables that change primarily over time, and where the differences between countries are less pronounced, e.g. the behavioral variables derived from event data. 

To classify which variables are static and dynamic we decomposed the extent to which they vary between countries and within countries, and consider those that vary more between countries than over time to be static. The Appendix \ref{app:variance} contains further details on this method.

\begin{figure}[htbp] 
   \caption{Conditional hazard and risk evolution in Thailand, based on the public discontent model}
   \label{fig:thailand}
   \centering
   \includegraphics[width=6in]{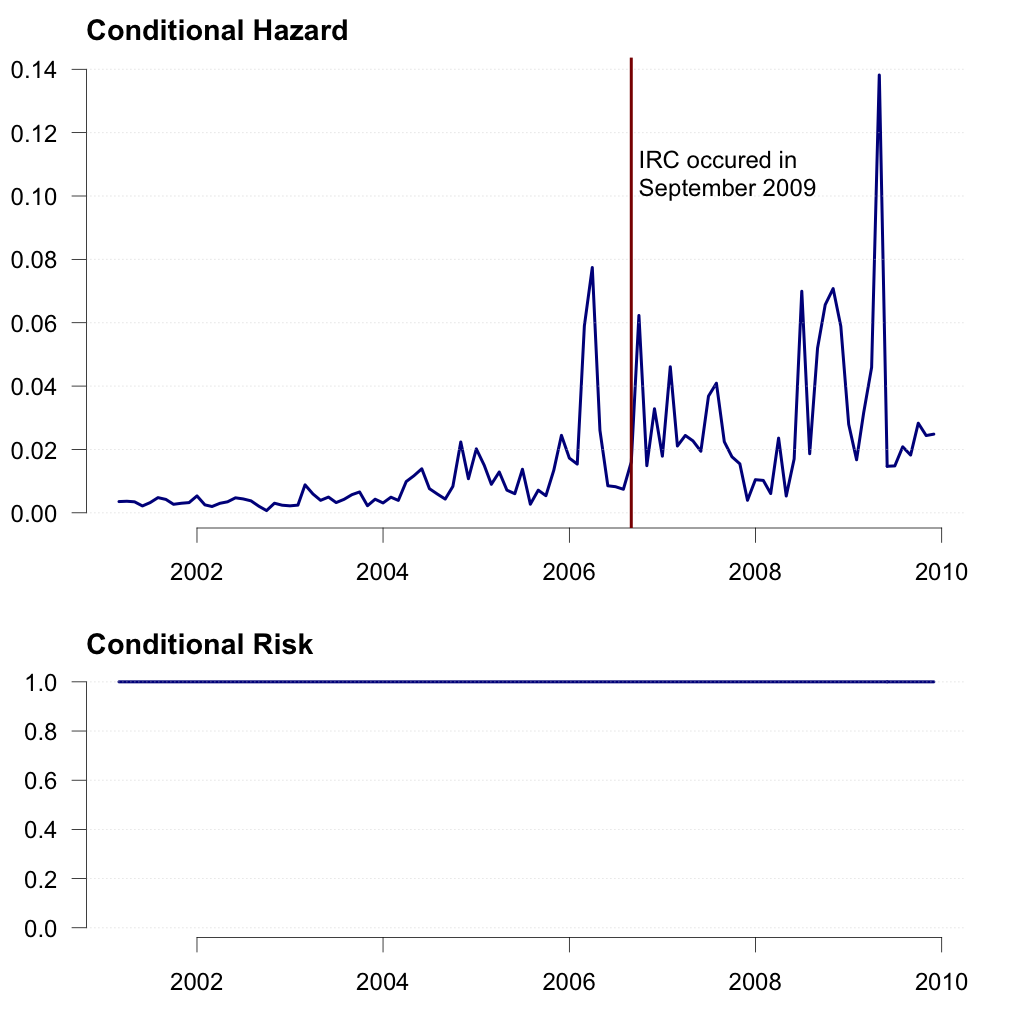} 
\end{figure}

The result of the variable selection scheme is that risk estimates for particular countries and with particular models are fairly consistent over time. Figure \ref{fig:thailand} shows the conditional risk and hazard estimates for Thailand generated from the public discontent model, which is discussed in further detail below. If we look at the risk estimates in bottom panel, they are constant over time, with a value of 1. This is in part because the inputs, like broadband users per 100, change little over time, and in part because Thailand is already at an extreme value for the risk input variables. The conditional hazard estimates in the top panel on the other hand fluctuate quite a bit. Recall that the conditional hazard is a function of the risk and unconditional hazard estimates. Since the former does not change, as we see in the panel below, these fluctuations are entirely due to the duration equation and further due to changes in the inputs to the duration equation. The duration equation thus captures the timing in this case.

\subsection{Ensemble model averaging}

The final piece in our design uses ensemble Bayesian model averaging to accommodate the heterogeneity in our dependent variable, ILC, and the heterogeneity in plausible mechanisms that can explain it. The concept of ensemble forecasting builds on the basic notion that
combining multiple points of view leads to a more accurate picture of
reality \citep[c.f.,][]{Surowiecki:2004}.  Among the more famous
demonstrations of this phenomenon was a competition to guess the
weight of an ox at the West of England Fat Stock and Poultry
Exhibition.  \citet{Galton:1907} famously demonstrated that, while
individual entrants were often wildly inaccurate, aggregating these into
an  average resulted in a
remarkably accurate estimate.\footnote{Draws on \citet{montegomery:etal:2014}.}

Ensembles have their modern birth in weather forecasting.  It was noticed  that
European weather forecasting was about 50\% more accurate than in the United States. A substantial
research project, funded by the Office of Naval Research, brought together atmospheric scientists,
psychologists, and statisticians to figure out why.  The result was ensemble Bayesian model averaging applied to weather forecasting, which averaged together different forecasts from different sensors, weighting them by how well they preformed in the past.  This approach is now in widespread use in the meteorological community, as well as in the arena of private weather forecasting, where there is a substantial derivatives market that has emerged.\footnote{An example of a public domain ensemble site is at \url{http://www.atmos.washington.edu/~ens/uwme.cgi}. Despite forecast, the European are still ahead: \url{http://news.nationalgeographic.com/news/2013/03/130307-weather-snowstorm-wrong-forecast-meteorology-world-europe-science/}.}

In recent years, the advantages of ensembles have come to play a
particularly prominent role in the machine-learning and nonparametric
statistics community \citep{Hastie:2009}. A wide range of approaches,
including neural nets, additive regression trees, and K nearest
neighbors, fall under the general umbrella of ensemble approaches.  Of
particular relevance is the success of boosting \citep{Freund:1997,
  Friedman:2001}, bagging \citep{Breiman:1996}, random forests
\citep{Breiman:2001}, and related techniques
\citep[e.g.,][]{Chipman:2010} to aggregate so-called ``weak
learners.''  These approaches to classification and prediction have
been advertised as the ``best off-the-shelf classifier[s] in the
world'', and are equally powerful in prediction
tasks.

While the advantages of collating information from multiple sources
are manifold, it is nevertheless false to assume that more is always
better.  Not all guesses are equally
informative, and naive approaches to collating forecasts risks
overvaluing wild guesses and undervaluing unusual forecasts that are
nonetheless sometimes correct.  The particular ensemble method we are
extending is ensemble Bayesian model averaging (EBMA). First proposed
by \citet{Raftery:2005}, EBMA pools forecasts as a weighted
combination of predictive PDFs.  Rather than selecting some ``best
model,'' EBMA collects \textit{all} of the insights from multiple
forecasting efforts in a coherent manner via statistical post
processing.  The weight assigned to each component forecast reflects
both its past predictive accuracy and its uniqueness (i.e., the degree
to which it makes predictions different from other component models).
It is important to emphasize that the intent of the EBMA approach is not to find the best 
model, but rather to find the right combination of models that will provide the best
overall predictions of some quantity of interest.

Assume the researcher is interested in predicting event
$\mathbf{y}^{t^*}$ for some future time period $t^\ast \in T^\ast$,
which we term the test period below. In addition, we have a number of
different out-of-sample forecasts for similar events $\mathbf{y}^t$ in
some past period $t \in T$, which we term the calibration period. The
different predictions were generated from $K$ forecasting models or
teams, $M_1, M_2, \ldots, M_K$.  These predictions might originate from the insights and intuitions of
individual subject-experts, traditional statistical models, non-linear
classification trees, neural networks, agent based models, or anything
in between.  Indeed, there is no restriction at all on the kind of
forecasting method that can be incorporated into the ensemble, so long
as it offers a prediction for a sufficiently large subset of the
calibration sample. Indeed, we use this approach with split-population, duration models.

For each forecast there is a prior
probability distribution $M_k\sim \pi(M_k)$ and the PDF for
$\mathbf{y}^t$ is denoted $p(\mathbf{y}^t|M_k)$. The
predictive PDF for the quantity of interest is
$p(\mathbf{y}^{t^*}|M_k)$, the conditional probability for each model
is given as 
\begin{eqnarray*}
p(M_k|\mathbf{y}^t) = p(\mathbf{y}^t|M_k)\pi(M_k)/ \underset{k=1}{\overset{K}{\sum}} p(\mathbf{y}^t|M_k)\pi(M_k), 
\end{eqnarray*}
and the marginal predictive PDF is 
\begin{eqnarray*}
p(\mathbf{y}^{t^*}) = \underset{k=1}{\overset{K}{\sum}} p(\mathbf{y}^{t^*}|M_k)p(M_k|\mathbf{y}^{t}).
\end{eqnarray*}
The prediction via EBMA is thus a weighted average of the component PDFs, and the weight for each model is based on its predictive performance on past observations in period $T$.

Our use employs specific models of themes for irregular leadership change, and averages over each of these themes. Returning to Figure \ref{fig:design}, the next partition in our data, from January 2010 to April 2012, forms the calibration data that is used in estimating the optimal way to average the different input models into one ensemble model. 

\subsection{Summary}

Our modeling process is as follows: we (1) create data on ILCs from March 2001 to March 2014, but with duration components informed by ILCs going back to 1955; (2) take a subset of this period, up to December 2009, to train several thematic models that each capture a difference facet of ILCs; (3) combine the individual thematic models in one ensemble model using EBMA and the period from January 2010 to April 2012 for calibration; (4) evaluate the resulting ensemble model predictions out-of-sample using our test data from May 2012 to March 2014; and finally (5) use the March 2014 data to create 6-month forecasts.

%
%
\section{Model results and forecasts}

We first present the individual thematic models, along with a discussion of their estimation results. We then turn to our discussion of the ensemble model and, finally, deliver the actual 2014 forecasts that are generated.

\subsection{Candidate models}

We included a total of 7 thematic split-population duration regression models in our final ensemble. Each model captures a different mechanism related to irregular leadership changes. We also explore how these different themes work in a temporal setting, whereby slower moving, structural variables enable us to best construct the risk equation, and more dynamic, quickly changing variables are utilized for the creation of hazard equations. We present these model specifications and results for each.

\subsubsection{Leader Characteristics}

Drawing on the literature on leadership tenure \citep{de2005logic, acemoglu2006persistence, svolik2012}, we examine a model that captures the leaders' individual characteristics as well as internal regime cooperation. The literature on leadership survival focuses on a leaders' ability to consolidate power over time, but also considers that as a leader consolidates power, they are more likely to create discontent among those who are not politically represented by the regime. The risk equation thus includes a count of the months a leader has been in power. To capture the legitimacy of a leader and by association his or her government, we include two further variables in the risk equation that indicate whether the current leader of a state entered power through irregular means or by foreign imposition. The thought is that leaders who entered through illegitimate, irregular means might themselves be more likely to suffer the same fate. 

The duration equation uses the material behavior of dissidents, whether cooperative or conflictual, to capture the timing. Since the type of state in which a leader has entered power irregularly is unlikely to be a well established regime with strong rule of law and civil society, we use material rather than verbal actions to model the timing of an ILC against illegitimate leaders (e.g. consider Afghanistan's Karzai or Michel Djotodia's tumultuous reign in the Central African Republic). 


\begin{table}[ht]
\centering
\begin{tabular}{rrr}
  \midrule
 & Estimate & p \\ 
  \midrule
(Dur. Intercept) & 6.37 & 0.00 \\ 
  log10(i.matl.conf.DIStGOV.l1 + 1) & -1.77 & 0.00 \\ 
  Material Cooperation Dissident $\rightarrow$ Government, logged and lagged & 0.98 & 0.33 \\ 
  \midrule
  Weibull Shape $\alpha$ & 0.51 & 0.00 \\ 
  \midrule
  (Risk Intercept) & -3.33 & 0.78 \\ 
  Leader with Irregular Entrance & -20.04 & 0.53 \\ 
  Leader who was Imposed by Foreign Actor & -51.75 & 0.52 \\ 
  Months in Power, logged & 24.05 & 0.54 \\ 
   \midrule
\end{tabular}
\caption{Leader characteristics model estimates} 
\label{tab:leader}
\end{table}

The results of this model are shown in Table \ref{tab:leader}. \emph{Positive} coefficients in the risk equation indicate higher probability of being in the risk set, while \emph{negative} coefficients in the duration equation indicate shortened survival time, and thus higher probability of ILC at any given time. Material conflict from dissident to government actors increases the probability of ILC with a statistically significant coefficient, as expected, while cooperation delays ILC. In the risk equation, leaders who enter power irregularly are themselves less likely to be at risk of irregular removal, and those imposed by foreign actors are less likely to experience ILC. The latter finding, although somewhat surprising, can be explained by the fact that the majority of foreign leader country-months in the data are leaders supported by the US: Hamid Karzai in Afghanistan, who has yet to lose power, Jay Garner and Paul Bremer in Iraq, and Sheikh Jabir al-Sabah of Kuwait, who is coded as being (re-)imposed by Coalition forces after the Gulf War. The only exception is Ahmad Tejan Kabbah, President of Sierra Leone from 1996 until his ouster during civil war in 1997, and again President from 1998 to 2008 after imposition through ECOWAS and later British intervention and support. This finding thus essentially indicates that leaders imposed by the US stay in power. Finally, leaders might be worse off the longer they are in power.

\subsubsection{Public Discontent}

As early as January 2013, Egypt's military voiced concerns over ongoing protests and their threat to recently elected Mohammed Morsi.\footnote{\href{http://www.washingtonpost.com/world/middle_east/egypts-military-chief-says-clashes-threaten-the-state/2013/01/29/8a8ee7ae-6a1b-11e2-ada3-d86a4806d5ee_story.html}{Washington Post, 29 January 2013} and \href{http://www.washingtonpost.com/world/middle_east/protesters-ransack-muslim-brotherhood-hq-demand-morsis-resignation/2013/07/01/f3f79698-e23c-11e2-a11e-c2ea876a8f30_story.html}{Washington Post, 1 July 2013}} Six months later the military caused and ILC by removing Morsi from power. The protests that had earlier ousted Mubarak and spurred the conflict that eventually ousted Qaddafi in Libya were hardly missed by the media. Many ILCs are preceded by visible intragovernment interactions as well as exchanges between dissidents and government actors. 

The public discontent model is focused on reports of such verbal interactions, as well as protests as an early warning indicator of ILCs. These are events such as speeches, expressions of support, threats, demands, and refusals involving actors associated with dissidents and government. These type of events often reveal actors' different preferences prior to when more substantive events take place. The model also includes counts of protests as a similar indicator. The final element in the duration equation is an indicator of verbal cooperation within government, primarily but not exclusively as an indicator of the health of civil-military relations. While even in autocracies there is level of competition between government agencies and sectors, the kind of military professionalization that is the norm in stable countries like the US includes strong proscriptions against public criticism of the civilian government. 

Since the level of public, verbal interactions in a society is related to media access and the resulting ability to voice demands, requests, etc., the model includes measures of internet users per 100 and cell subscribers per 100 people in the risk equation. For example, while a certain level of verbal conflict between dissidents and government might be normal in a relatively open country like the US where media access allows for easy fora to voice opposition, reports of verbal conflict at a similar level from a country with little modern information infrastructure are indicative of more significant interactions. 

Furthermore, there is research showing that mass communication through media like newspapers, radio, television, and the internet play an important role in the dynamics between regime opponents and supporters, which suggest that cell phone and internet usage are related to ILC through their impact of collective action as well. Unsurprisingly, many authoritarian governments implement censorship to control the information available to citizens. Recent work on China shows that censorship is targeted specifically to undermine collective action, rather than criticism of the regime \textit{per se} \citep{king2013censorship}. Yet, diffuse forms of mass communication, i.e. the internet and broad adoption of cell phones, are inherently more difficult to censor than traditional mass media like the press, radio, and television, and thus pose a new threat to regimes that may be proficient in censoring other forms of media \citep{edmond2013information}. 

Lastly, we include the fraction of excluded population in a country as a control since it might be correlated with the indicator of verbal cooperation within government: minority governments facing a large opposition have strong incentives to display unity.


\begin{table}[ht]
\centering
\begin{tabular}{rrr}
  \midrule
 & Estimate & p \\ 
  \midrule
(Dur. Intercept) & 5.88 & 0.00 \\ 
  Verbal Cooperation within Government, logged and lagged & 0.97 & 0.03 \\ 
  Verbal Conflict from Government $\rightarrow$ Dissident, logged and lagged & -1.17 & 0.16 \\ 
  Verbal Conflict from Dissident $\rightarrow$ Government, logged and lagged & -1.26 & 0.11 \\ 
  Number of Anti-Government Protests, logged and lagged & -0.98 & 0.04 \\ 
  \midrule
  Weibull Shape $\alpha$ & 0.29 & 0.03 \\ 
  \midrule
  (Risk Intercept) & 115.52 & 0.59 \\ 
  Internet Users, lagged & -42.95 & 0.61 \\ 
  Cell Phone Users, lagged & 32.31 & 0.62 \\ 
  Excluded population, logged and lagged & 22.94 & 0.78 \\ 
   \midrule
\end{tabular}
\caption{Public interactions model estimates} 
\label{tab:chatter}
\end{table}

Table \ref{tab:chatter} show the model estimation results. The effects in the duration equation are in keeping with what one might expect: verbal conflict from government to dissident actors and \textit{vice versa} hastens ILC, as do protests, while high verbal cooperation within the government, suggesting a united state, suppresses ILC. In the risk equation, the total number of internet users reduces risk, but a large number of cell phone users is associated with an increased probability of being in the risk set. Similarly, a large excluded population, indicative of an ethnic minority government, increases risk. None of the variables are statistically significant however.

\subsubsection{Global instability}

Our third model is loosely based on the main components of the \citep{goldstone2010global} model. In this paper, the authors use a conditional logistic regression model with four variables centered on a new regime classification scheme that allows them to identify partial democracies with factionalism to forecast global instability, which includes adverse leadership changes. Crucially, they use a case control method to deal with the rarity of positive events, where positive cases are matched to a random subset of negative control cases. 

Using their findings on the conditions that drive global instability, we have created a model loosely based on theirs, but necessarily different given our different modeling strategy and data resolution. In our version, the partial democracy with factionalism indicator did not perform as well as simply including the Polity participation of competitiveness variable (``PARCOMP''). The latter is described by the Polity project as capturing whether ``alternative preferences for policy and leadership can be pursued in the political arena.'' Our final version of the global instability model is shown below. Echoing the Goldstone approach, we include GDP and the percent of the population excluded from the political process into the risk equation. Then, to predict the timing of ILC, we include participation competitiveness, a measure of conflict within the four nearest neighbors, as well an indicator of female life expectancy at birth. 


\begin{table}[ht]
\centering
\begin{tabular}{rrr}
  \midrule
 & Estimate & p \\ 
  \midrule
(Dur. Intercept) & 3.44 & 0.00 \\ 
  Spatial Lag of Insurgency Events in Nearest 4 Neighbors, lagged & 32.36 & 0.75 \\ 
  Female Infant mortality, lagged & 0.04 & 0.03 \\ 
  Factionalism, lagged & -0.36 & 0.09 \\ 
  \midrule
  Weibull Shape $\alpha$ & 0.14 & 0.39 \\ 
  \midrule
  (Risk Intercept) & 7.59 & 0.00 \\ 
  GDP per capita, lagged & -1.73 & 0.01 \\ 
  Excluded Population, lagged & -1.05 & 0.64 \\ 
   \midrule
\end{tabular}
\caption{Global instability model estimates} 
\label{tab:pitf5}
\end{table}

Surprisingly, neighboring conflict delays ILC, and infant mortality has a significant but very small impact by delaying conflict. This may be due to the broader nature of our dependent variable. Factionalism on the other hand hastens ILC, as expected. In the risk equation, wealth and excluded population also have expected effects. 

\subsubsection{Protest}

Our fourth thematic model is entirely focused on protest. As is empirically evidenced in cases across the globe, civil resistance campaigns are an effective means for achieving leadership change. The literature on both coup-proofing \citep{quinlivan1999coup,pilster2011coup} and civil resistance campaigns \citep{chenoweth2011} describe a key force behind protest movements: their ability to influence the military. A pivotal movement in many civil resistance campaigns is the moment when state forces stop obeying orders from the head of state, and refuse to openly repress protestors. This model captures the basic intuition of this argument by including slower moving structural variables, such as low levels of domestic crises and military expenditure, into the risk equation. This model is structured by the argument that the least satisfied militaries will be most likely to resist commands to repress. In the duration equation we employ more variant covariates that account for protest and conflict in different forms: ethnic-religious violence, rebellion, protest events, and nearby rebellion events in other countries. 

\begin{table}[ht]
\centering
\begin{tabular}{rrr}
  \midrule
 & Estimate & p \\ 
  \midrule
(Dur. Intercept) & 6.10 & 0.00 \\ 
  N low-intensity conflictual deeds:ethnic groups \& government actors & 7.86 & 0.93 \\ 
  N low-intensity conflictual deeds: rebel groups \& government & 11.67 & 0.89 \\ 
  N protest events directed against all actors & -0.01 & 0.07 \\ 
  Low-intensity conflict in countries w/ similar political structure & -0.11 & 0.00 \\ 
  \midrule
  Weibull Shape $\alpha$ & 0.38 & 0.01 \\ 
  \midrule
  (Risk Intercept) & 0.91 & 0.20 \\ 
  N high-intensity conflictual deeds & 0.65 & 0.19 \\ 
  Military Spending, lagged and logged & 4.78 & 0.11 \\ 
   \midrule
\end{tabular}
\caption{Protest model estimates} 
\end{table}

The model results show that conflict does not have a straightforward relationship with ILC. While protests and low-intensity conflict in surrounding countries with similar political structures hasten ILC, low-intensity conflict, either with ethnic groups or other armed rebel groups, within a country seems to strengthen the regime, or at least reduce the chances of ILC. As we find with some of the other models, there appear to be conditions in which certain types of conflict are associated with increased government stability. Considering where these types of armed conflicts occur, i.e. mainly in Africa and south Asia, it seems that low levels of conflict can  be used by leaders who otherwise lack legitimacy to bolster their regimes.  

\subsubsection{Contagion}

Our last model captures the concept of conflict contagion. How and whether conflict is a phenomenon that is effected by contagion-like processes is an on-going theoretical and empirical question in political science \citep{saideman2012conflict}. 
To model the risk for successful contagion of mass protests or other conflict that may lead to an ILC, we include the country's Amnesty International Political Terror Scale value, which captures overall repressiveness, as well as opposition resistance, which counts the number of resistance events conducted by groups associated with the opposition, rebels, or insurgents in a country. The latter largely varies between rather than within countries, and we thus include it as a static variable.\footnote{See the Appendix \ref{app:variance} for details on how we classified static, structural variables versus dynamic variables.} These two variables are intended to capture the overall security climate in a country. To further refine the general risk posed in a repressive society with ongoing terror or political violence, which is captured by the first two variables, we include an indicator of temporally proximate elections. This variable identifies whether an election will occur in the near future or has occurred in the near past. 

Finally, we include the country's population size as an indicator of society's inertia and resistance to outside influences. For example, we would expect that a small country is on average more sensitive to events in it's neighboring countries than a country with a large population, in which attention is necessarily more domestically oriented. 

With this equation for risk, we then use two spatial weights of opposition resistance and state repression in neighboring countries to model the timing until contagion, and hence increased chance of ILC, occurs. The spatial weights are constructed based on the distance between the geographic centers of countries, and thus countries that are close have more influence than countries further away. With the inclusion of population in the risk equation, our overall model is thus an approximation of a gravity model principal, where influence is weighted by size and distance. 

\begin{table}[ht]
\centering
\begin{tabular}{rrr}
  \midrule
 & Estimate & p \\ 
  \midrule
(Dur. Intercept) & 8.22 & 0.00 \\ 
  Number of events of resistance in neighboring countries & -5.63 & 0.07 \\ 
  Number of repressive events in neighboring countries & 3.00 & 0.26 \\ 
  \midrule
  Weibull Shape $\alpha$ & 0.02 & 0.86 \\ 
  \midrule
  (Risk Intercept) & -1.67 & 0.43 \\ 
  Amnesty International Terror Scale, lagged & 1.53 & 0.00 \\ 
  Proximity of the Next Election & -0.62 & 0.37 \\ 
  Number of Acts of Resistance, lagged and logged & 1.56 & 0.04 \\ 
  Population, lagged & -0.63 & 0.38 \\ 
   \midrule
\end{tabular}
\caption{Contagion model estimates} 
\label{tab:contagion}
\end{table}

Table \ref{tab:contagion} shows the contagion model results. Resistance in neighboring countries increases the risk of ILC, while repression reduces it. This indicates a demonstration effect might be at work: successful resistance close by can inspire home-grown efforts, while repression deters them by indicating lower chances of success. 

\subsubsection{Internal conflict}

The internal conflict model uses GDP per capita, the proximity of the next national election, and the level of Autocracy in the country as general indicators of risk, while focusing on intra-governmental conflict and the widespread use of cell technology as duration triggers.  Intra-governmental tensions, protests to the government, and the use of cell phones are taken to interact to influence the duration of leadership tenure and the likelihood of an irregular transfer.  First order components of this interaction also are included in the duration equation, but the second order-interactions (e.g. the two-way interactions) are excluded as they cause instabilities in the likelihood. Results of the estimation are given in Table~\ref{tab:internal}.

\begin{table}[ht]
\centering
\begin{tabular}{rrr}
  \midrule
 & Estimate & p \\ 
  \midrule
(Dur. Intercept) & 7.12 & 0.00 \\ 
  Internal tension $\times$ Protests $\times$ Cell Phones, logged and lagged & 0.33 & 0.04 \\ 
  Intra-Governmental Tension, logged and lagged & -0.16 & 0.61 \\ 
  Anti-governmental Protest, logged and lagged & -1.45 & 0.00 \\ 
  Cell Phones, logged and lagged & -0.21 & 0.30 \\ 
  \midrule
  Weibull Shape $\alpha$ & 0.40 & 0.01 \\ 
  \midrule
  (Risk Intercept) & 2.33 & 0.99 \\ 
  GDP per capita, logged and lagged & -14.95 & 0.87 \\ 
  Proximity of Next Election, lagged & 19.12 & 0.91 \\ 
  Level of Autocracy, lagged & -7.36 & 0.89 \\ 
   \midrule
\end{tabular}
\caption{Internal conflict model estimates} 
\label{tab:internal}
\end{table}

The interaction term prevents us from directly interpreting the individual effects in the duration equation. When assuming that one of the other terms is 0 however, we can say that each term by itself is negatively related to regime stability. For example, in the absence of intra-governmental tension or anti-government protests, cell phone subscriber rates are associated with higher chances of ILC. The positive coefficient on the interaction term however indicates that in combination, the effects of the 3 variables are somewhat attenuated. This suggests a saturation effect, i.e. that at high levels of tension and protest in a country with high cell phone usage, increasing any particular factor has a diminishing impact on the chances of ILC.

\subsubsection{Finanical}

This model assumes that financial instability may unseat leaders who are already in a precarious situation. The baseline risk is determined by GDP per capita, as a measure of general prosperity, the looming presence of the next election, as well as the size of the country as measured by population. In addition, it includes the Amnesty assessment of terrorism (stability) and the degree of anti-government.  If a country is in the high risk set, it is the degree of inflation, as measured by consumer prices, and the health of the countries international financial reserves (taken from the IMF's IFS statistics) that affect most directly the duration of leadership. 

\begin{table}[ht]
\centering
\begin{tabular}{rrr}
  \midrule
 & Estimate & p \\ 
  \midrule
(Dur. Intercept) & 24.03 & 0.00 \\ 
  Inflation, via CPI, logged & -5.93 & 0.00 \\ 
  International Reserves, logged & 0.45 & 0.00 \\ 
  \midrule
  Weibull Shape $\alpha$ & 0.19 & 0.22 \\ 
  \midrule
  (Risk Intercept) & -0.28 & 0.97 \\ 
  GDP per capita, logged and lagged & -3.06 & 0.12 \\ 
  Amnesty International Terror Scale, lagged & 2.40 & 0.02 \\ 
  Proximity of Next Election, logged and lagged & 1.20 & 0.70 \\ 
  Opposition Resistance, logged and lagged & 8.66 & 0.18 \\ 
  Population, logged and lagged & 3.94 & 0.15 \\ 
   \midrule
\end{tabular}
\caption{Financial Stability model estimates} 
\label{tab:fin}
\end{table}

The model estimates in Table \ref{tab:fin} show that inflation can be an indicator of impending ILC. Conversely, large international reserves can serve as a buffer. Repression is costly, especially considering that the kind of social upheaval that might lead states to repress make traditional revenue collection difficult. International reserves might facilitate repression of opponents in regimes in which repression otherwise might not be financially feasible. 

In the absence of open dissent, large international reserves also provide rulers with the capacity to buy off dissent. This is consistent with research on the role of food prices, and state's ability to subsidize them, on regime instability in the Middle East \citep{lagi2011food, sadiki2000popular}.

\section{Ensemble and 2014 forecasts}

\subsection{Model weights and goodness of fit}

The seven thematic models above are aggregated into a final ensemble model using weights that are based in part on the overall fit of the input models, as well as the uniqueness of their predictions. Table \ref{tab:ensemble} shows the ensemble weights, as well as the component model fits in the training, calibration, and test periods. The model weights are applied to thematic model predictions that are rescaled to reduce prediction bias, and thus the ensemble models is in essence a weighted combination of rescaled thematic model predictions. 

We show two statistics here for model fit. The Appendix contains further statistics and a more detailed discussion of model fit in sparse binary data and how to interpret related statistics. First, we include the classification accuracy, calculated as the number of true positives and true negatives over the total number of cases. In other words, accuracy measures how many country-months a model ``gets right.'' While simple, classification accuracy is potentially misleading in sparse data like ours, since it is a lot easier to predict 0's than 1's. Furthermore, the accuracy and other statistics like precision and recall depend on a particular cut point value to bin the model predictions into discrete 0 or 1 values. Changing this cut point will change the accuracy, precision, recall, false positive rates, etc. For example, if we increase the cut point, our model will be more conservative with fewer false positives, but also fewer true positives. 

The AUC, or area under the ROC curve, is insensitive to the specific choice of a cut point. It measures the area under the receiver operating characteristic (ROC) curve, which itself is a plot of the model tradeoff between correctly predicting positive events and falsely predicting negative (non-)events as the cut point changes. A random guess on average has an AUC value of 0.5, and 1 indicates a perfect model that can classify ILC occurrences perfectly. Higher values of AUC are thus better. The AUC can also be interpreted as the chances that the model will rank a case randomly picked from the set of positive cases higher than a case randomly picked from the set of negative cases. In this interpretation it is easier to see why a random guess will have an AUC of 0.5: without knowledge of which case is more likely to be positive and negative, it is essentially a coin flip.


\begin{table}[ht]
\centering
\begin{tabular}{lrrrrrrr}
  \toprule
   &&\multicolumn{2}{c}{Training}&\multicolumn{2}{c}{Calibration}&\multicolumn{2}{c}{Test} \\
   \cmidrule{3-4}  \cmidrule{5-6}  \cmidrule{7-8} 
Model & W & AUC & Accuracy & AUC  & Accuracy  & AUC  & Accuracy  \\ 
  \midrule
Ensemble &  & 0.845 & 0.997 & 0.890 & 0.993 & 0.839 & 0.992 \\ 
  Leader char. & 0.006 & 0.803 & 0.987 & 0.816 & 0.986 & 0.776 & 0.983 \\ 
  Public disc. & 0.007 & 0.848 & 0.995 & 0.832 & 0.992 & 0.820 & 0.994 \\ 
  Global Inst. & 0.006 & 0.779 & 0.964 & 0.717 & 0.998 & 0.726 & 0.998 \\ 
  Protest & 0.006 & 0.746 & 0.985 & 0.658 & 0.997 & 0.752 & 0.995 \\ 
  Contagion & 0.681 & 0.824 & 0.998 & 0.839 & 0.992 & 0.814 & 0.885 \\ 
  Internal conflict & 0.290 & 0.788 & 0.997 & 0.829 & 0.993 & 0.743 & 0.992 \\ 
  Financial & 0.006 & 0.782 & 0.999 & 0.789 & 0.953 & 0.787 & 0.979 \\ 
   \bottomrule
\end{tabular}
\caption{Ensemble model, monthly observations} 
\label{tab:ensemble}
\end{table}

The first column in the table, W, gives the weight assigned to each model's prediction when aggregating to the ensemble model. Larger weights indicate that that thematic model plays a more important role in the ensemble, while low weights indicate a model's impact is negligible. The contagion model's predictions have the largest weight, about two-thirds in the final ensemble model. The internal conflict  model  also make a significant contribution. The remaining  models have a marginal impact on the ensemble.

We include training fit for completeness, and it is apparent that individually, the public discontent and contagion models have the highest fit in terms of AUC, while the accuracies are fairly similar. The model predictions for the calibration period are the actual basis for the EBMA model weights and ancillary parameters. Looking at the summary fit statistics, the overall trends in fit are similar, with a spread of AUC's from 0.717 to 0.839 for the thematic models. Note that while the EBMA weights are roughly related for a model's fit in the calibration period, the EBMA also weighs models with unique predictions more and down-weighs models whose predictions are correlated. Calibration fit is thus not linearly related to the EBMA weights. 

The last two columns in Table \ref{tab:ensemble} shows the fit of the theme and ensemble models in the test period. These summary statistics capture how well the model fits out-of-sample, and are thus indicative of the model's forecast performance. The ensemble model, as one can see, has a AUC of 0.839. The AUC indicates that if we were to randomly pick two country months, one with a ILC, the other without an event, the ensemble model on average has a 0.839 chance of assigning a larger ILC probability to the positive case. To put this into context, this means that the ensemble model has a 0.839 chance, on average, of assigning a higher probability to one of the 8 country-months in our test period during which an ILC occurred over the 3,764 country-months in which an ILC did not occur.  

Some of the component models have a higher AUC value in the test period than the ensemble model, even though the ensemble model has the highest AUC in the test period. This is due to the sparsity of the dependent variables. Even slight changes in the number of ILC's can have a big impact on model performance. 

\subsection{Fit in annualized data}

\begin{table}[ht]
\centering
\begin{tabular}{lrrrrrrr}
  \toprule
   &&\multicolumn{2}{c}{Training}&\multicolumn{2}{c}{Calibration}&\multicolumn{2}{c}{Test} \\
   \cmidrule{3-4}  \cmidrule{5-6}  \cmidrule{7-8} 
Model & W & AUC & Accuracy & AUC  & Accuracy  & AUC & Accuracy  \\  
  \midrule
Ensemble &  & 0.944 & 0.978 & 0.906 & 0.986 & 0.959 & 0.982 \\ 
  Leader char. & 0.006 & 0.835 & 0.981 & 0.775 & 0.982 & 0.750 & 0.982 \\ 
  Public disc. & 0.007 & 0.927 & 0.982 & 0.845 & 0.978 & 0.929 & 0.986 \\ 
  Global Inst. & 0.006 & 0.920 & 0.910 & 0.902 & 0.978 & 0.919 & 0.982 \\ 
  Protest & 0.006 & 0.924 & 0.981 & 0.923 & 0.986 & 0.986 & 0.994 \\ 
  Contagion & 0.681 & 0.930 & 0.980 & 0.848 & 0.949 & 0.911 & 0.974 \\ 
  Internal conflict & 0.290 & 0.937 & 0.982 & 0.891 & 0.970 & 0.950 & 0.990 \\ 
  Financial & 0.006 & 0.881 & 0.982 & 0.896 & 0.974 & 0.876 & 0.982 \\ 
   \bottomrule
\end{tabular}
\caption{Model fit, annualized data} 
\label{tab:fit.country.year}
\end{table}

Since previous PITF modeling efforts have been based on country-years, it is misleading to compare our model's country-month fit directly to those previous efforts. Increasing the resolution in the data invariably makes it more sparse, thus increasing the difficulty of the modeling problem. To sidestep this problem we present model fit for annualized data in Table \ref{tab:fit.country.year}. To annualize the data, we marked a country-year as experiencing a failure if at least one of it's constituent country-months had an event, and we aggregated monthly to yearly predictions by calculating the probably of any event with the formula $p_{c, yearly} = 1 - \prod (1 - p_{c, monthly})$. The ensemble and thematic model accuracies are almost entirely above 98\%, and several of the AUC values are above 0.9, indicating a good fit.\footnote{The accuracy and AUC can be lower than in the monthly data because of different in the optimal cut point for the monthly and yearly versions of a model.}

To derive a quantity that is easier to interpret in light of the 6-month forecasts we present below, we also calculated how many of the model positive predictions occur within 6-months of an ILC. Using the optimal threshold presented above, while the recall rates are fairly low at $\sim$25\%, the revised precision estimate with a 6-month window is $\sim$50\%. This provides evidence that while our model might not be able to anticipate some ILCs, we can be somewhat confident that a significant fraction of the model's predictions will come true.

The Appendix contains a more extensive discussion of model fit, including how we derived the estimate in the preceding paragraph. 


\subsection{Recall and precision}

\subsection{2014 Forecasts}

Using the ensemble model and data from March 2014 we calculate forecasts for the probability of ILC over the next 6 months, i.e. for the period from April to September 2014. We aggregate the monthly forecasts produced by this model to an overall probability of ILC anytime during this time period,\footnote{Using $p^* = 1 - \prod_{t=1}^6 (1 - p_t)$, where $p_t$ is the forecast $t$ months ahead.} and Table \ref{tab:forecast} show the 20 highest forecasts that result. 

\begin{table}[ht]
\centering
\begin{tabular}{lr}
  \midrule
Country & Probability \\ 
  \midrule
Ukraine & 0.28 \\ 
  Bosnia and Herzegovina & 0.19 \\ 
  Yemen & 0.10 \\ 
  Egypt & 0.07 \\ 
  Thailand & 0.06 \\ 
  Guinea & 0.05 \\ 
  India & 0.04 \\ 
  Turkey & 0.04 \\ 
  Libya & 0.03 \\ 
  Central African Republic & 0.03 \\  \midrule
  Honduras & 0.03 \\ 
  Guinea-Bissau & 0.03 \\ 
  Greece & 0.03 \\ 
  Russia & 0.03 \\ 
  Afghanistan & 0.03 \\ 
  Romania & 0.03 \\ 
  Cote d'Ivoire & 0.03 \\ 
  Bangladesh & 0.03 \\ 
  Mali & 0.02 \\ 
  Israel & 0.02 \\ 
   \midrule
\end{tabular}
\caption{Top 20 forecasts for ILC between April and September 2014 (6 months) using March 2014 data} 
\label{tab:forecast}
\end{table}

\begin{figure}[htbp] 
   \centering
   \caption{Forecasts for ILC between April and September 2014 (6 months) using March 2014 data}
   \label{fig:6months}
   \includegraphics[width=6.4in]{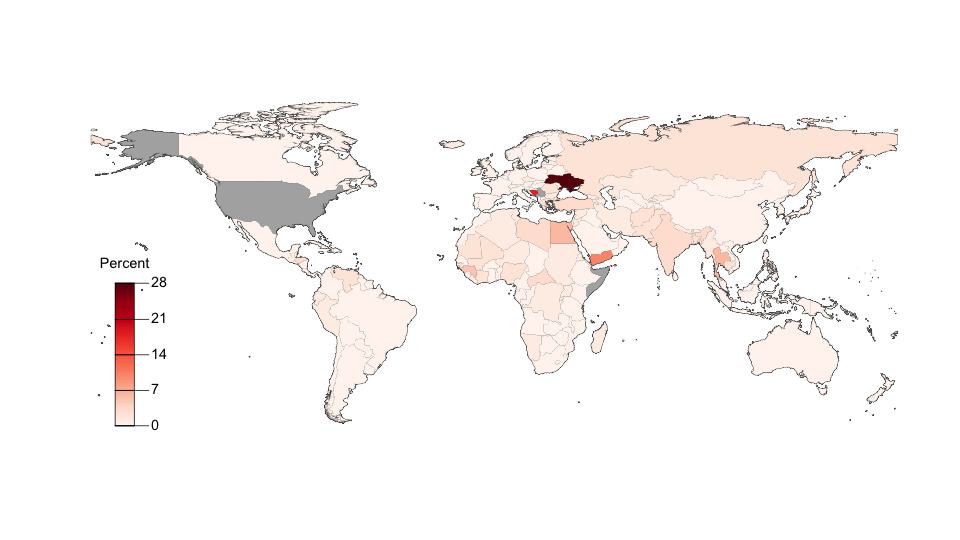} 
\end{figure}

We enumerate a brief summary of some countries at the top of the risk predictions for the six months beginning in April, 2014.
\begin{enumerate}
\item[] {\bf Ukraine} lost the Crimea to Russia this spring, but the protests in the winter of 2013 continued into the new year and large scale, violent protests occurred in the middle of February, 2014 which were in part a response to the so-called Anti-Protest laws enacted in the previous month.  By the end of February, Parliament essentially ousted the President, and scheduled a May 25 election.  This created a succession crisis in which the deposed president, Yanukovich and his supporters in Russia began to create a larger conflict in the Ukraine. This Russian involvement spawned further local conflict within Ukraine owing to the many pro-Russian Russians living in the Ukraine. The ILC that occurred in February falls in our test period, and the model predictions increase ten-fold from the level in November up until the occurrence of ILC in February. However, a continuation of conflict within government and separatist activities in the east, in combination with a new government, place Ukraine at the top of our model predictions. 

\item[] {\bf Bosnia and Herzegovina} has been the site of substantial anti-government protests in early 2014, the so-called Bosnian Spring.  These were organized in large part because of the frailty of the economy, the high level of unemployment, and the non-payment of pensions.  Spreading from Tuzla, to Mostar, on to Sarajevo and even Banja Luka, these anti-government protests have been widespread. Prime Minister Vjekoslav Bevanda has minimized the protests. However, as the leader of an increasingly weak central government, the greatest instability may  reside in the locally governed regions. 

\item[] {\bf Yemen} is the site of some protests against the government, accompanied by the presence of a very powerful Al-Qaeda army. That, accompanied by a set of rulers widely reported to be very corrupt, creates a very unstable situation.  Yemen is in the throes of (another) reorganization in which central authority is likely to devolve to regional ruling coalitions.

\item[] {\bf Egypt} sees an outbreak of protests and violence every year in February to celebrate the resignation of Mubarak and the start of the so-called revolution in the early spring of 2011. There is continued and sporadic protests throughout the country over the year. In mid-2013 protests led to the ousting of Morsi by the Military via an coup d'\'etat.  The crisis over the succession continues, although in mid-2013 Mansour was appointed as acting president. In early 2014 a new constitution was overwhelmingly ratified by Egyptian voters, even though roughly $2/3$ of the potential voters avoided participating.  Tension remains high and the legitimacy and popularity of the current regime is tenuous at best..

\item[] {\bf Thailand} has been a puzzling cauldron of political conflict for over a decade. Thaksin Shinawatra was overthrown by a coup d'\'etat in the fall of 2006, and the resulting junta instituted martial law and forbade many political activities until mid-2007. Things continued to be contentious and violent, but by mid-2011 things had calmed down and Yingluck Shinawatra handily won the election. Toward the end of 2013, protests heated up quite a bit as did demands for the resignation of Yingluck.  Schedule elections were not held in 2014, as the listing of the candidates outraged anti-government forces.  As this was written, on May 7th, the (Supreme) Constitutional court ruled that Yingluck had abused power and was to be removed from the Prime Ministership.  It is unclear how this will turn out, except to note that this will constitute another irregular leadership change in early 2014.

\item[] {\bf Guinea} was the site of protests over the elections in May.  The response by the government led to more widespread protests, which evolved into conflicts between long-standing opponents.  An election was scheduled for May 12, but as yet it is unclear whether this will occur.  What is clear is that there will be widespread protests as well as government responses. How exactly it will turn out is unclear, but the likelihood of an irregular leadership change is not zero.

\item[] {\bf India} continues to govern without coalitions, but the existence of many factions.  Currently in the throes of a general election to finish later in May. The United Progressive Alliance received the barest of margins to form a government without a coalition in 2012 (33.2\%), but there are many relatively sizable Indian political parties.

\item[] In {\bf Turkey,} the Gezi protests included about 2.5 million protesters in roughly 80 cities and resulted in the arrest of almost 5000. It is estimated that 4000 protesters and over 600 police were injured. Four were killed.\footnote{These data were taken from the Turkish Human Rights Association report, entitled: 
{\em Gezi Parki Direni\c{s}i Ve Sonrasinda Ya\c{s}ananlara Ili\c{s}kin De\u{g}erlendirme Raporu}, 3/4/2014.} Widespread corruption, repression, and protest do not bode well for a smooth leadership transition in the 10 August 2014 elections, especially since the current Prime Minister Ergo\u{d}an is at the eye of the storm.

\item[]  {\bf Libya}, while having elections for parliament in July 2012, there is still simmering conflict in Libya. As yet, no new constitution has been drafted. Numerous militia are a powerful force in many cities. Clashes between Gaddafi loyalists and the central government continued into 2014.  In the forecast period (at the time of writing) many reports of violence have emerged, even one which was described as a failed coup. In July, clashes re-emerged and escalated to such a level that the Tripoli airport was closed and the US Embassy was evacuated.

\item[] {\bf The Central African Republic} 2013 coup d'\'etat lead to a great deal of instability, and as in January 2014, the Prime Minister resigned leaving the country in a state of crisis. As a poor country with a transitional, interim President, CAR is unstable.  In February, the UN deployed troops in the country, sending French and African Union troops on site to keep the peace. However, this peace is fragile and the potential survival of the current authorities is highly problematic.

%


\end{enumerate}

%
%
\section{Discussion}

We used new, temporally disaggregated data that included behavior variables derived from event reports; we  employed split-population duration and ensemble modeling approaches to examine irregular leadership changes over the period from 1955 to the present.  To our knowledge each of these aspects is novel in the study of leadership change. In so doing we also developed a suite of new empirical models, focusing on leader characteristics, public discontent, protest behavior, contagion, intra-governmental conflict, as well as financial and economic instability. We combined this with an implementation of the model developed for Phase V of the PITF, though our implementation was based on monthly, not yearly data.  In addition to these novel aspects, we then combined the forecasts of each of these empirical models by using ensemble Bayesian model averaging to produce a single probability estimate that benefits from the so-called ``wisdom of the crowds.''   Along the way, we also had to update the dependent variable over the past 2 and one-half years.  

Many months pass in each country without an irregular leadership change. Irregular leadership changes are rare.  Our model combines a base probability of whether a country is at risk for such a change, as well as an assessment of triggers that might push an at risk country toward having an irregular leadership change. Still, we are looking for needles in a haystack. The suite of models we developed are examined in historical, training data and were evaluated in test data that were not used in their construction. In both of these contexts, the models are accurate and well-calibrated. Finally, we use an ensemble combination of these models that is weighted to produce six-month forecasts of the conditional hazard over the period from April 2014 until September 2014.  These predictions are discussed above, but in summary
seem plausible.  Indeed, two of the top twenty forecasts are the Ukraine and Thailand, both currently in the throes of transition crises.  The probabilities for these twenty predictions are mostly around $0.04$ or lower and most of the rest of the PITF countries examined have probabilities which are even smaller.  

Yet, it should be remembered that for every twenty estimates that there is a $0.05$ chance of rain, for a properly calibrated model, you should expect it to rain at least once.

%
%
\newpage \clearpage
\bibliographystyle{apsr}
\bibliography{arch_refs}
\addcontentsline{toc}{section}{References}

%
%
\clearpage
\appendix
\section{Analyses of the Boundaries}
\setcounter{table}{0}
\renewcommand{\thetable}{A\arabic{table}}
\subsection{List of Irregular Regime Transitions}



 

\begin{table}[ht] 
 \caption{List of Irregular Exits and Entries, Sorted by Date}
\label{tab:list}
\centering
\scalebox{.68}{
\begin{tabular}{lllcccrr}
  \toprule
\sc Country & \sc Date & \sc Leader & \shortstack{\sc Entry\\\sc Irregular} & \shortstack{\sc Exit\\ \sc Irregular} &  \shortstack{\sc ILC} & \sc TTF & \shortstack{\sc Months \\ \sc in Power} \\ 
  \midrule
Afghanistan & 2001-11 & Mullah Omar &   1 &   0 &   1 &  62 & 62 \\ 
  Madagascar & 2002-07 & Ratsiraka &   1 &   1 &   1 &  65 & 329 \\ 
  Central African Republic & 2003-03 & Patasse &   1 &   0 &   1 & 113 & 258 \\ 
  Liberia & 2003-08 & Taylor &   1 &   0 &   1 &  72 & 155 \\ 
  Guinea-Bissau & 2003-09 & Kumba Iala &   1 &   1 &   1 &  43 & 52 \\ 
  Georgia & 2003-11 & Shevardnadze &   1 &   0 &   1 & 140 & 142 \\ 
  Haiti & 2004-02 & Aristide &   1 &   0 &   1 &  36 & 149 \\ 
  Ecuador & 2005-04 & Lucio Gutierrez &   1 &   0 &   1 &  27 & 63 \\ 
  Kyrgyzstan & 2005-04 & Akayev &   1 &   0 &   1 & 174 & 164 \\ 
  Mauritania & 2005-08 & Sidi Ahmed Taya &   1 &   1 &   1 & 248 & 248 \\ 
  Nepal & 2006-04 & Gyanendra &   1 &   0 &   1 &  14 & 616 \\ 
  Solomon Is. & 2006-05 & Rini &   1 &   0 &   1 &   1 & 71 \\ 
  Thailand & 2006-09 & Thaksin Shinawatra &   1 &   0 &   1 &  67 & 187 \\ 
  Fiji & 2006-12 & Laisenia Qarase &   1 &   0 &   1 &  77 & 77 \\ 
  Bangladesh & 2007-01 & Iajuddin &   1 &   0 &   1 &   3 & 193 \\ 
  Fiji & 2007-01 & Senilagakali &   0 &   1 &   1 &   1 & 1 \\ 
  Georgia & 2007-11 & Saakashvili &   1 &   0 &   1 &  46 & 48 \\ 
  Lebanon & 2008-05 & Siniora &   1 &   0 &   1 &   6 & 641 \\ 
  Mauritania & 2008-08 & Ould Cheikh Abdellahi &   1 &   1 &   1 &  16 & 36 \\ 
  Guinea & 2008-12 & Conte &   0 &   1 &   1 & 296 & 296 \\ 
  Guinea-Bissau & 2009-03 & Vieira &   1 &   0 &   1 &  41 & 66 \\ 
  Latvia & 2009-03 & Godmanis &   1 &   0 &   1 &  15 & 210 \\ 
  Madagascar & 2009-03 & Marc Ravalomanana &   1 &   1 &   1 &  80 & 80 \\ 
  Honduras & 2009-06 & Zelaya &   1 &   1 &   1 &  41 & 370 \\ 
  Guinea & 2009-12 & Dadis Camara &   1 &   1 &   1 &  12 & 12 \\ 
  Niger & 2010-02 & Mamadou &   1 &   1 &   1 & 122 & 130 \\ 
  Kyrgyzstan & 2010-04 & Bakiyev &   1 &   0 &   1 &  60 & 60 \\ 
  Tunisia & 2011-01 & Zine Al-Abidine Ben Ali &   1 &   0 &   1 & 278 & 278 \\ 
  Egypt & 2011-02 & Mubarak &   1 &   0 &   1 & 352 & 352 \\ 
  Cote d'Ivoire & 2011-04 & Laurent Gbagbo &   1 &   1 &   1 & 126 & 126 \\ 
  Libya & 2011-08 & Qaddafi &   1 &   1 &   1 & 503 & 503 \\ 
  Maldives & 2012-02 & Nasheed &   1 &   0 &   1 &  39 & 561 \\ 
  Romania & 2012-02 & Emil Boc &   1 &   0 &   1 &  38 & 266 \\ 
  Yemen & 2012-02 & Saleh al-Hashidi &   1 &   0 &   1 & 403 & 404 \\ 
  Mali & 2012-03 & Amadou Toure &   1 &   1 &   1 & 117 & 252 \\ 
  Guinea-Bissau & 2012-04 & Raimundo Pereira &   1 &   1 &   1 &   3 & 37 \\ 
  Mali & 2012-04 & Amadou Sanogo &   1 &   1 &   1 &   1 & 1 \\ 
  Greece & 2012-05 & Loukas Papadimos &   1 &   1 &   1 &   6 & 462 \\ 
  Guinea-Bissau & 2012-05 & Mamadu Ture Kuruma &   1 &   1 &   1 &   1 & 1 \\ 
  Libya & 2012-08 & Mustafa Jalil &   1 &   0 &   1 &  12 & 12 \\ 
  Bulgaria & 2013-03 & Boyko Borisov &   1 &   0 &   1 &  44 & 699 \\ 
  Central African Republic & 2013-03 & Francois Bozize &   1 &   1 &   1 & 120 & 120 \\ 
  Egypt & 2013-07 & Morsi &   1 &   1 &   1 &  13 & 29 \\ 
  Central African Republic & 2014-01 & Djotodia &   1 &   1 &   1 &  10 & 10 \\ 
  Ukraine & 2014-02 & Yanukovych &   1 &   1 &   1 &  48 & 267 \\ 
   \bottomrule 
\multicolumn{7}{p{6in}}{{\sc TTF} $=$ time to failure in months.} \\
\end{tabular}
}
\end{table}x

\subsection{ILC, irregular entry and exit, and coups}
\label{app:coupsILC}

Our dependent variable, irregular leadership change, is a composite of two variables from the Archigos dataset which code whether a leader entered power through irregular means or left power through irregular means \citep{archigos2009codebook, goemans2009introducing}. ``Irregular'' denotes entry or exit that does not follow the established rules or conventions of a state for governing leader transitions, which can range from democratic elections to one-party rule to hereditary monarchy. 

For example, a military leader who stages a coup and assumes power would be considered to have had an irregular entry. A rebel leader who overthrows the central government would also have an irregular entry. A party leader who becomes prime minister after winning an election would have a regular entry, as would a prince who assumes power after the king or queen's abdication in a monarchy, or a Politburo member who assumes the most powerful positions in a one-party state after the retirement of the preceding leader. 

On the other hand, a leader who loses power as a result of a successful coup against him or her would lose power in an irregular manner, like Morsi in Egypt in 2013, as would one overthrown by a successful rebellion, like Gaddafi in Libya in 2011. A leader who resigns in the face of mass protests would lose power in an irregular manner, e.g. Yanukovich in Ukraine in 2014. Leaders who lose elections, who lose parliamentary confidence, or who retire in the absence of clear, direct pressure lose power in a regular manner. Leaders who retire due to ill-health or die in office also are not considered to have left power irregularly. 

As these examples make clear, there is conceptual and empirical overlap between irregular entries and exits. A coup almost always implies the irregular exit of the leader who was overthrown, and the irregular entry of whoever staged the coup. Table \ref{tab:entryexit} shows the overlap between irregular entries and exits in our data. Rows depict regular and irregular exits, respectively, while columns depict regular and irregular entries, respectively. There are a total of 43 irregular exits and 21 irregular entries, of which 19 overlap.

Most irregular entries also entail an irregular exit, for situations like the coup example above. The two exceptions are December 2008 Guinea and January 2007 Fiji. In Guinea, the military staged a coup after the death of long-time ruler President Cont\'{e}. In Fiji, the caretaker prime minister Senilagakali, who had been installed after the 2006 coup, resigned during the restoration of Ratu Josefa Iloilo as President by the military leader Bainimarama. Bainimarama was then appointed interim primer minister the next day. Senilagakali's resignation is not coded as irregular in the original Archigos data, although it may be hard to untangle the complicated political situation during this time period. Aside from these 2 exceptions, irregular entries are a subset of irregular exits, with most corresponding to leaders who took power through successful coups. 

Approximately half of irregular exits are not followed by irregular entries of the succeeding leaders. Most of these cases capture successful mass protests or armed rebellions which lead to the irregular exit of a current leader, and where the successors are designated either by an established line of succession or through elections following a short interim government. 

\begin{table}[ht]
\centering
\begin{tabular}{rrr}
  \hline
 & 0 & 1 \\ 
  \hline
0 & 26671 &   2 \\ 
  1 &  24 &  19 \\ 
   \hline
\end{tabular}
\caption{Irregular exits and entries} 
\label{tab:entryexit}
\end{table}

Our dependent variable, ILC, is simply a combination of irregular entries and exits, coded as occurring when either an irregular entry or exit or both occur in a country in a given month. As the preceding discussion shows, there is partial overlap between these two inputs, specifically in that irregular entries are largely a subset of irregular exits.

Table \ref{tab:coups} shows the overlap between ILC and coups, using the \citet{powell2011global} list of coups. Values of 1 and 2 in the rows depict attempted and successful coups. As one would expect, non of the attempted but unsuccessful coups are captured in ILC, and 16 of the 17 successful coups are. The sole exception is Togo in February 2006, a case similar to the Guinea coup discussed above. The President, Gnassingb\'{e} Eyad\'{e}ma, died in office, after which the military installed his son, Faure Eyad\'{e}ma, as President rather than the speaker of the legislature, who was next in line for succession. Gnassingb\'{e} resigned shortly under international pressure, but was elect as President in April of the same year. His assumption of power is not coded as irregular by the original Archigos data. Coups are thus largely a subset of ILC. 

\begin{table}[ht]
\centering
\begin{tabular}{rrr}
  \hline
 & 0 & 1 \\ 
  \hline
0 & 26654 &  29 \\ 
  1 &  16 &   0 \\ 
  2 &   1 &  16 \\ 
   \hline
\end{tabular}
\caption{P\&T Coups and ILC} 
\label{tab:coups}
\end{table}

\subsection{Variable variance decomposition} \label{app:variance}

Variables by definition change, some more so, some less so, the extent of which can be captured by a variable's variance, which is defined as the overall deviation of individual values from an overall average. It's a truism that one cannot predict something, i.e. a dynamic outcome like ILC, with nothing, i.e. a variable that does not change. Having variables with large variance is thus a necessary, although not sufficient, condition for creating models that fit and predict well. 

In the context of country-month panel data, and with the task of predicting outcomes that occur at different times in different countries, the variance of variables can also be further decomposed into changes that occur between countries and changes that occur over time. At the extremes are variables which only vary between countries, but not over time, like land area, and variables that change only over time, not between countries, like global oil prices. Most variables obviously have some combination of these two variance structures. 

Most of the variables that are traditionally used in IR, like regime types and economic indicators, vary more between countries than they do over time within any particular country. For example, while GDP, to the great concern of most people, changes in countries over time, it does so at fairly moderate and somewhat predictable rates, with annual growth rates of $\sim2\pm6\%$.\footnote{Using 2011 and 2012 GDP for all countries from the World Development Indicators; approximate 95\% margin.} On the other hand, the spread of countries' GDP around the global mean is much larger, with a standard deviation of 15,400 US dollars around a mean of 10,684 USD.

One obvious implication is that we cannot predict the timing of events within a country using variables that do not change that much within any given country over time. Many of the variables used in country-year studies of conflict as a result break down at the country-month level, unless it is possible to meaningfully disaggregate them to a monthly level. 

Furthermore, the interpretability of split-population duration model estimates is more straightforward if the risk estimates are particular to countries and do not change dramatically within a given country over time. This requires identifying \emph{static} variables that mainly vary between countries, but not within, to serve as inputs for the risk equation. 

To identify static variables which, for any given country, do not change much over time, we decomposed the variance for the covariates into variance between countries and variance within countries. A relatively simple way to do this is by decomposing the total sum of squares (SS). The total SS is the sum of squared differences between a variables's values at time $t$ for country $c$ and the overall mean of that variable:
\begin{eqnarray*}
SS_{total} = \sum (x_{c, t} - \bar{x})^2
\end{eqnarray*}
High values of $SS_{total}$ essentially indicate a larger spread around the overall mean value, while low values indicate less divergence from the mean, and hence less variance. To decompose the total SS we further calculate the SS between countries and the SS within countries using the average value of a variable within each country, $\bar{x}_c$:
\begin{eqnarray*}
SS_{total} & = & SS_{between} + SS_{within} \\
SS_{between} & = & \sum (\bar{x}_c - \bar{x})^2 \\
SS_{within} & = & \sum (x_{c, t} - \bar{x}_c)^2 
\end{eqnarray*}
The SS between countries is thus a function of the spread of each country's average value of a variable around the overall mean, while the SS within countries is a function of the spread of the values for a country around that country's average value. The former indicates large differences between countries, the latter large differences within countries. 

\begin{figure}[htbp] 
   \centering
      \caption{Variance decomposition}
   \includegraphics[width=6in]{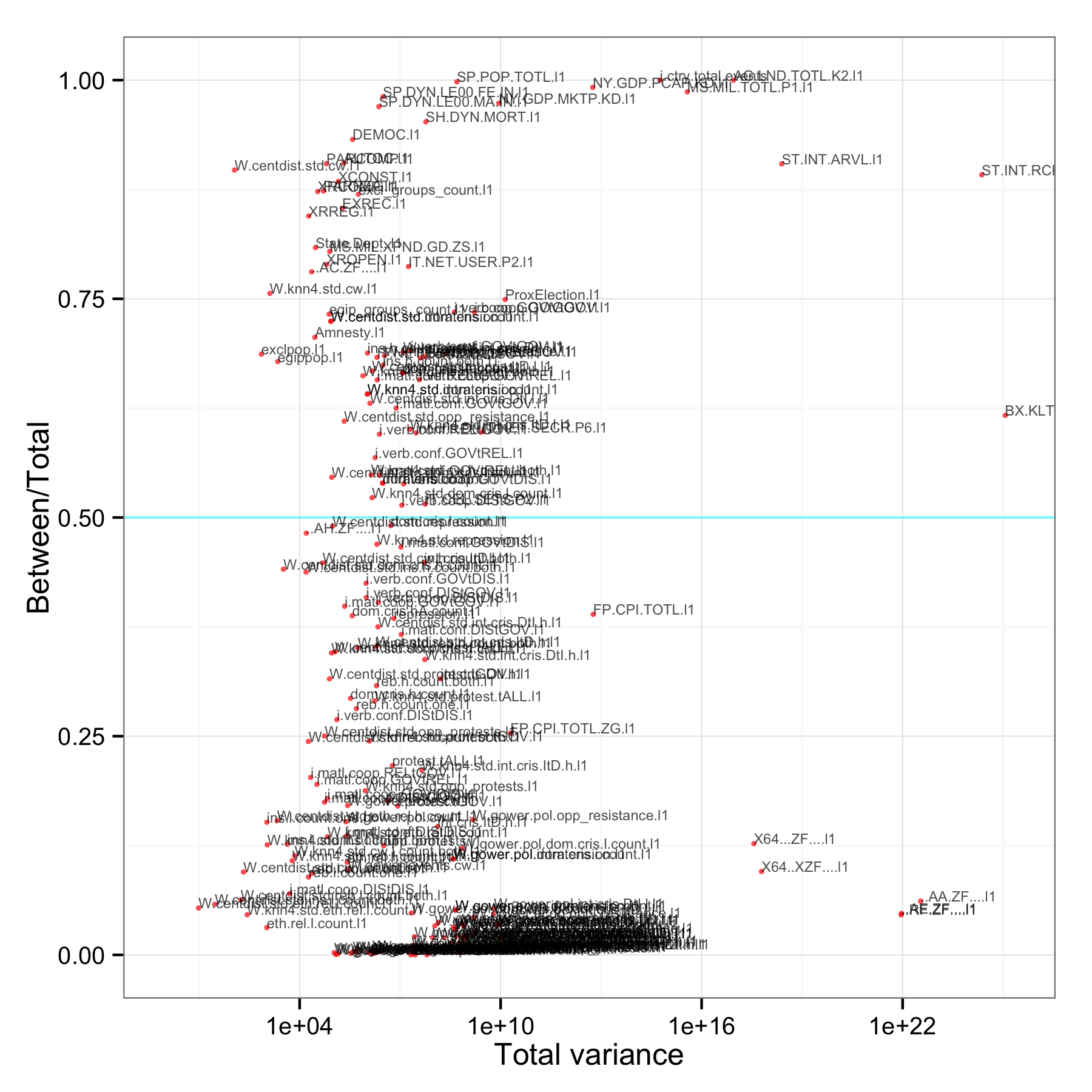} 
   \label{fig:all}
\end{figure}

There is a large space of possible combinations between total variance and variance within and between countries for any given variable. Figure \ref{fig:all} visualizes these combinations for all variables in our dataset. The $x-$axis gives the overall variance of a variable, with higher values to the right. The $y-$axis shows the portion of a variable's total variance that is between, rather than within, countries. Variables in the top half of the figure thus vary largely between countries, and variables in the bottom half vary largely within countries over time. We use this distinction for classifying static from dynamic variables. 

\begin{figure}[htbp] 
   \centering
   \begin{subfigure}[b]{0.48\textwidth}
   	\includegraphics[width=\textwidth]{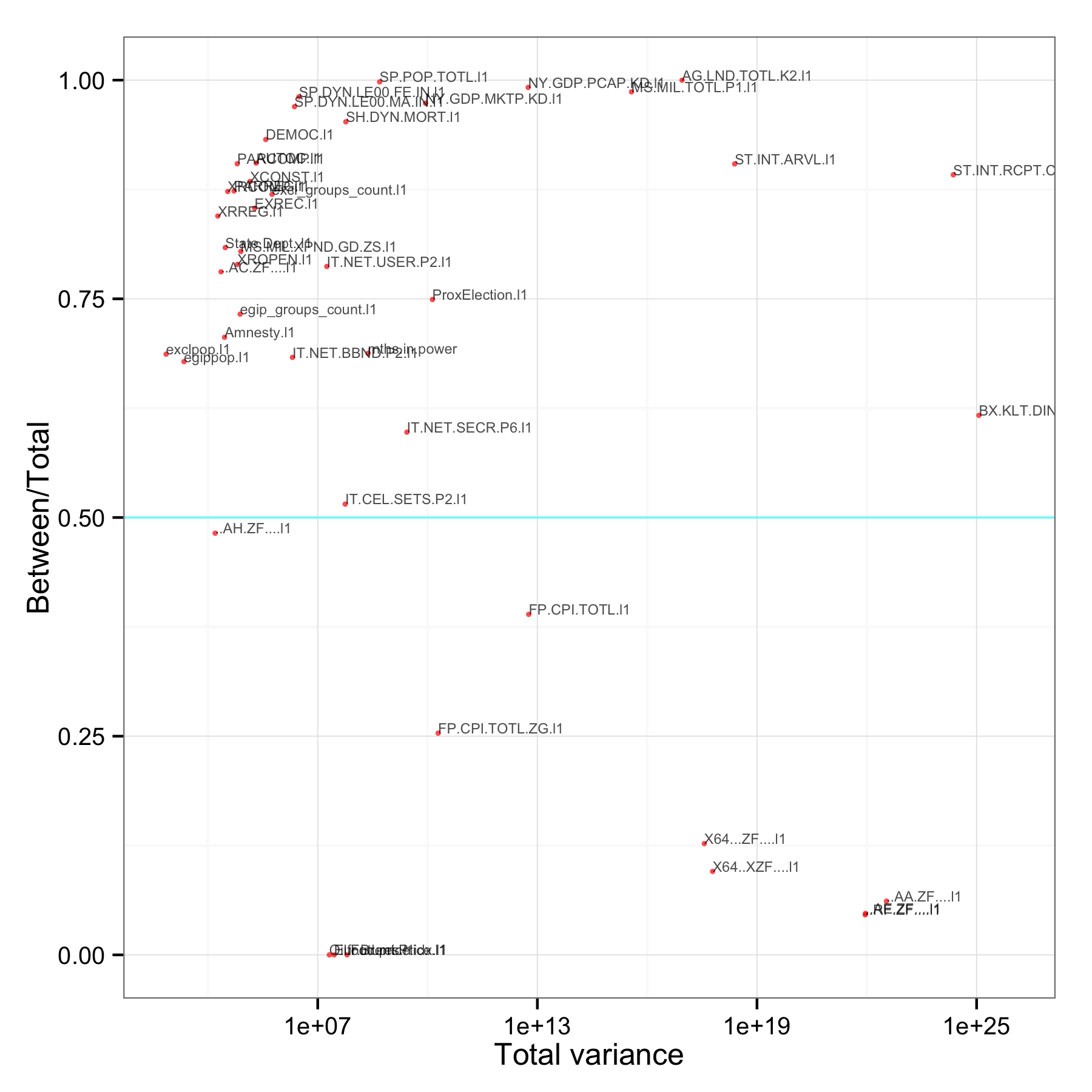}
	\caption{Structural}
   \end{subfigure}
   \begin{subfigure}[b]{0.48\textwidth}
   	\includegraphics[width=\textwidth]{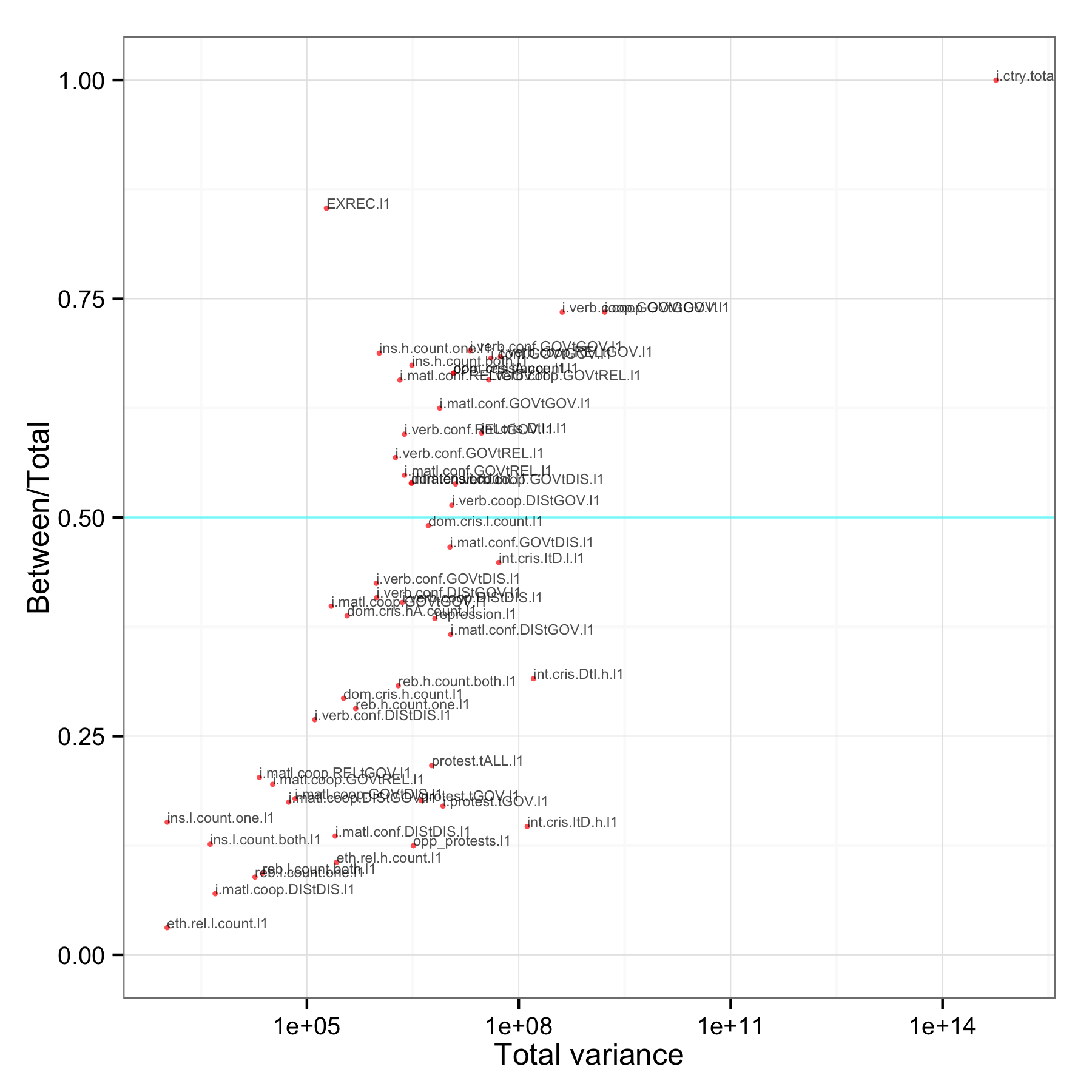}
	\caption{Event aggregations}
   \end{subfigure}
   \begin{subfigure}[b]{0.48\textwidth}
   	\includegraphics[width=\textwidth]{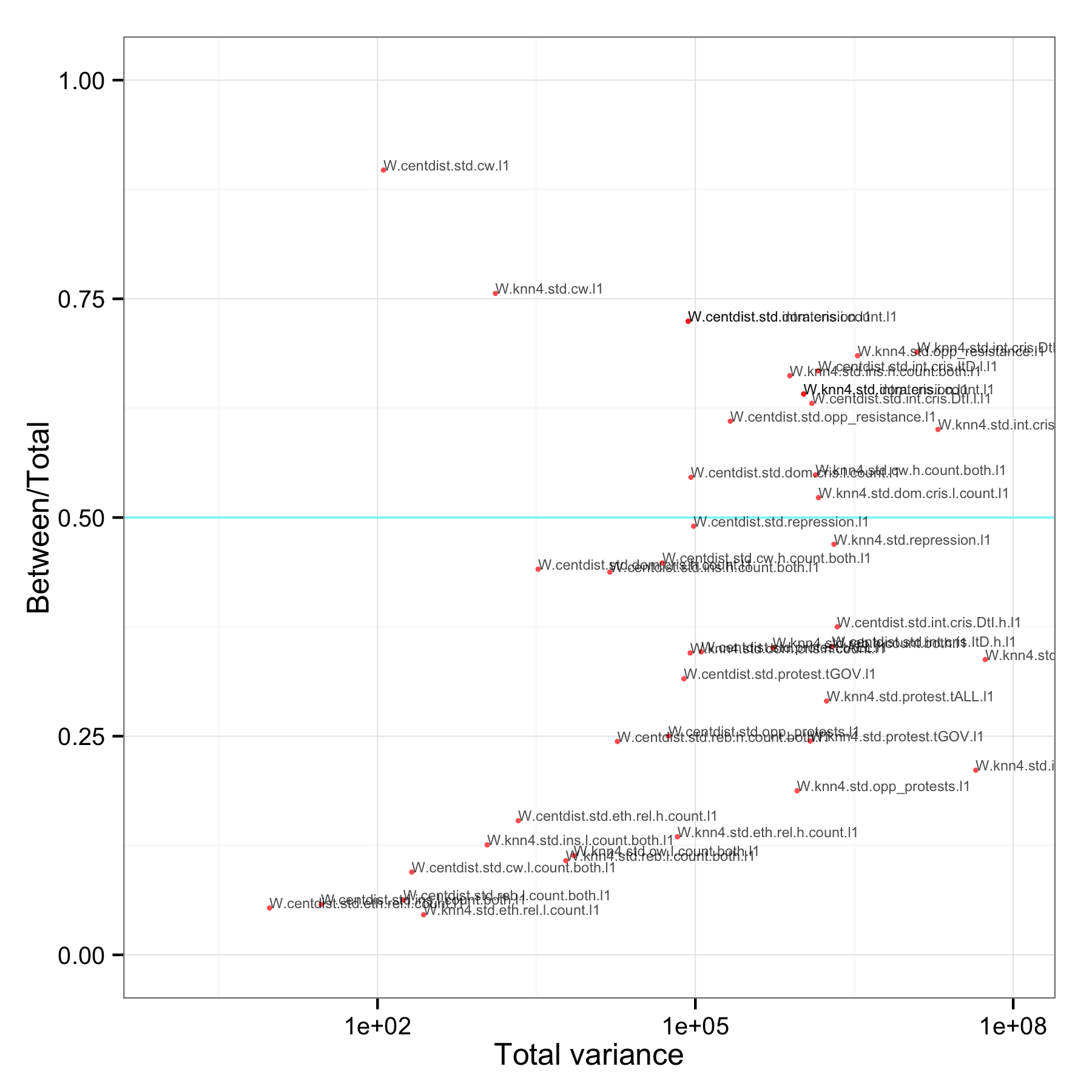}
	\caption{Spatial weights}
   \end{subfigure}
   \begin{subfigure}[b]{0.48\textwidth}
   	\includegraphics[width=\textwidth]{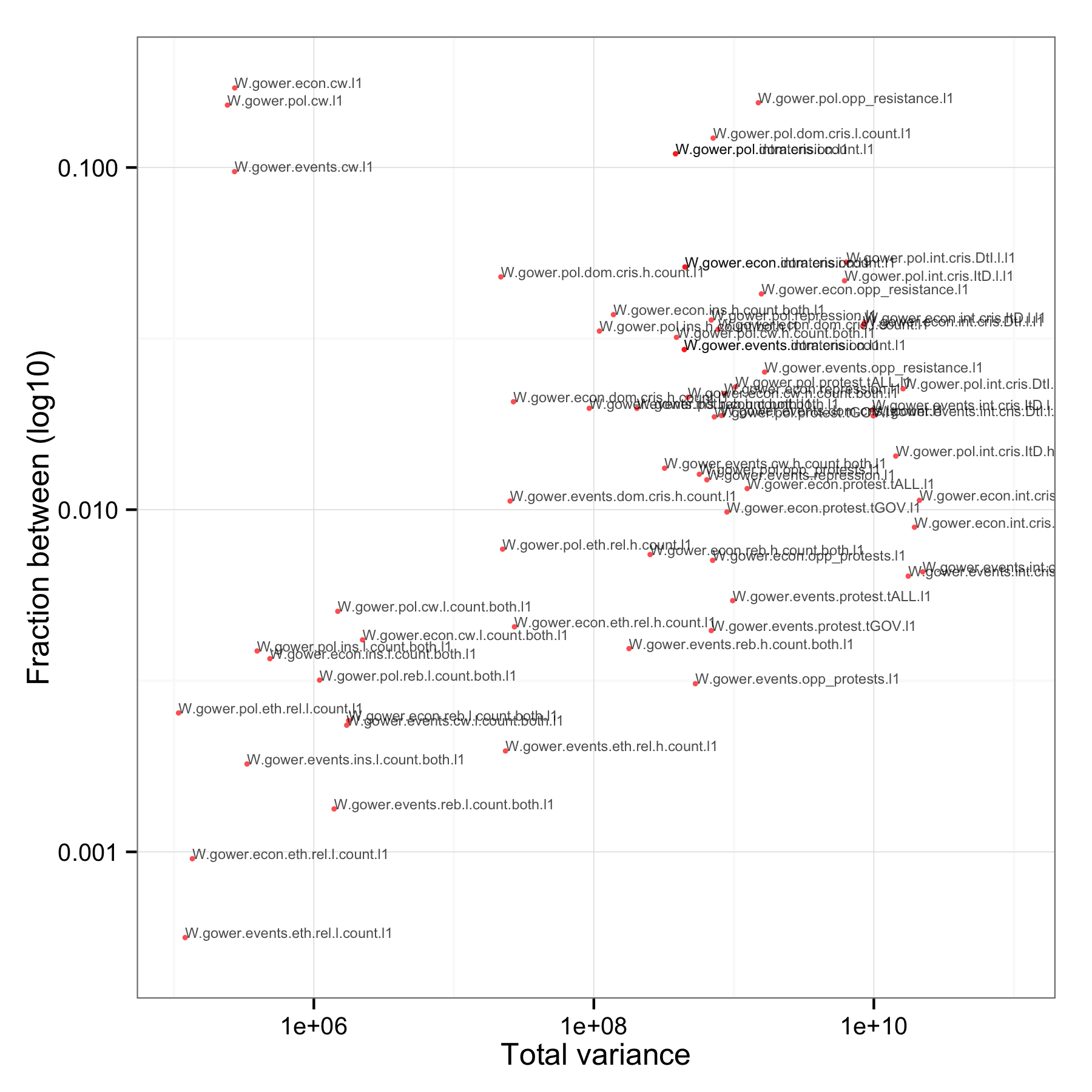}
	\caption{Spatial weights, Gower only}
   \end{subfigure}
   \caption{Variance decomposition}
   \label{fig:fcast-sep}
\end{figure}

The panels in Figure \ref{fig:fcast-sep} further break down variance by sub-groups of variables.

\subsection{ICEWS event data descriptives}

The Integrated Crisis and Early Warning System (ICEWS) project generates event data as part of a larger effort to forecast different types of political violence. Originally the data covered the time period from 2001 on, with monthly updates that are produced with a delay of a few days, but more recently they have also been back coded to 1991. The events are machine coded from a corpus of stories, with an approach and software similar to that used by GDELT and earlier instantiations like the Kansas Event Dataset \citep[KEDS;][]{schrodt2006twenty}, and in contrast to event data efforts that rely on human coding, like ACLED \citep{raleigh2010introducing} and UCDP GED \citep{sundberg2013introducing}. Since the data originate in a defense and intelligence community, rather than academic, project, they are so far not publicly available. As a result there are no published evaluations of the ICEWS event data, although internally evaluations of the coding accuracy have been conducted both by us and other participants in the larger project.

\begin{figure}[h!]
\caption{ICEWS events by month}
\centering
\includegraphics[width=6.5in]{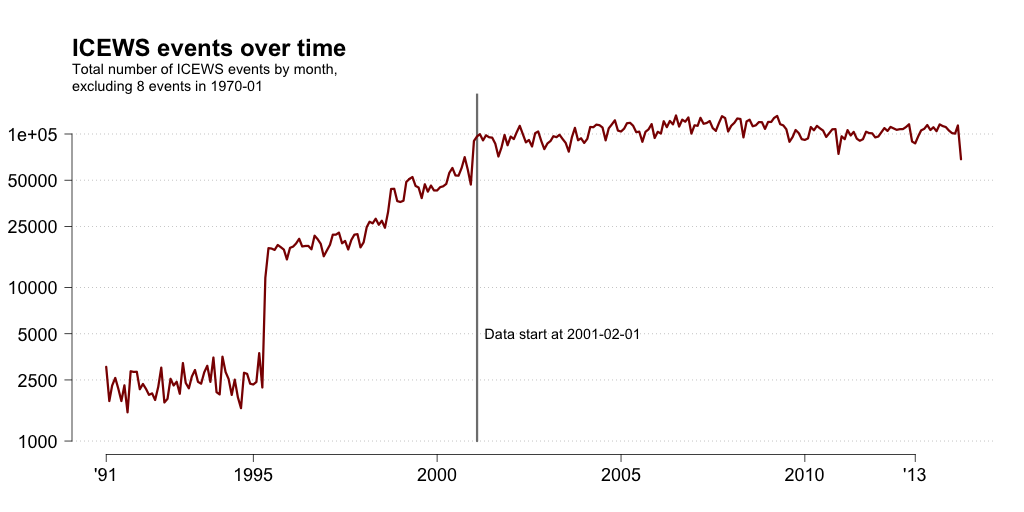}
\label{fig:icews-time}
\end{figure}

A fundamental problem with event data, machine- or human-coded, is their relationship to media bias, which can distort the relationship between some set of events in the world and media stories about them, and the overall volume of media coverage, which can distort the apparent number of coded events by introducing false positives and duplicate coding of events. If we assume that the number of events in the world is more or less constant over a time span covering a few years, then we would expect a relatively constant number of events that are coded by a machine or human-based effort. Figure \ref{fig:icews-time} shows the total number of events per month for the time period covered by the latest version of ICEWS, January 1st 1991 through December 31st 2013. Note that the $y$-scale is logarithmic, and that changes in the line are therefore more abrupt than they appear. There are a total of 18.5 million events over this whole period. The number of events is relatively constant from 2001 on, and in fact this period contains the majority, 16 million, of total events. This is the starting point for the data used in the ICEWS forecasting effort, and we use it as a starting point for our data as well. 

The remaining 2.5 million events are scattered through the period from 1991 to 2000, with a sharp discontinuity in 1996 and exponential increase thereafter. The discontinuity is likely the result of a technical change in the story corpus data source.  On the other hand, the exponential increase beginning in 1996 is the result of the dramatic increase in the corpus of stories available as a basis for coding from the mid-1990's on, due to the increasing proliferation of the internet and availability of online media. This exponential growth in the underlying story corpus, from which event data are coded, is a general trend and one of the major problems for machine-coded event data. As the line show, there is successful effort to correct for the growth in stories in the ICEWS data from 2001 on. The number of events in the GDELT event data on the other hand match the exponential growth in stories, thus leading to efforts to normalize the data before use in studies that cover more than a few years.

\begin{figure}[h!]
\caption{ICEWS events per capita for 2001--2013 by country}
\centering
\includegraphics[width=6.5in]{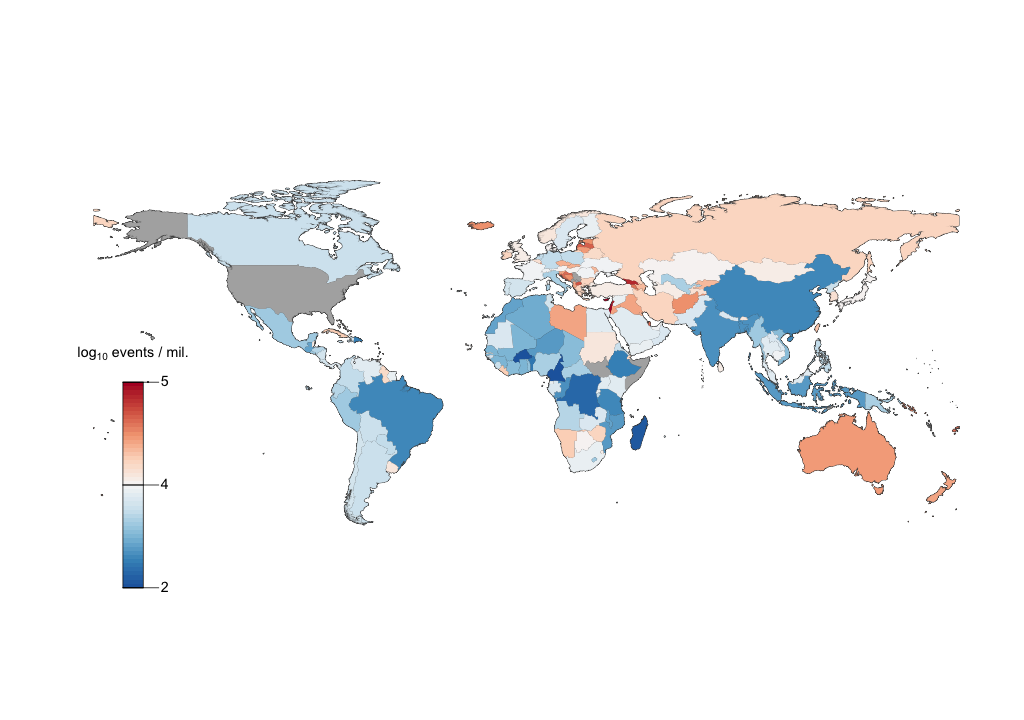}
\label{fig:icews-map}
\end{figure}

One of the other concerns with event data is systematic media bias. Without a ``ground truth'' reference set of actual events, we have to speculate about the nature of this bias. Assuming that events are mainly the function of population interactions, we would expect that the number of events per capita is relatively constant across countries. Figure \ref{fig:icews-map} show the log of events per million for the countries included in our data. Per capita counts range from under 300 events per million for Cameron and Madagascar or 500 per million for China to more than 40,000 events per million in Israel and Lebanon. Brazil, Sub-saharan Africa, India, China, and Indonesia stand out as having abnormally low per capita counts.

What are the implications of this apparent reporting variation? Considering that most irregular leadership exits occur in places that may have under-reporting, it undermines the predictive usefulness of behavioral indicators. Suppose high protest activity really is related to irregular exit. If protests in Western Europe and the US, where irregular leadership changes are rare, are reported at much higher levels than protests in Africa, the data might drown out the relationship between protests and our outcome of interest. There have been various efforts to normalize the GDELT data, e.g. by using country and year or month totals for event counts. We have similarly tried to normalize by country totals, without success. 

\subsection{Left-censoring}

The split-population model that we use assumes that we accurately know the time at which the last previous  ILC occurred. For example, Thailand's Thaksin Shinawatra was removed by a military coup in September 2006. The last previous  ILC in Thailand occurred in February 1991, when Chatichai Choonhavan was also ousted by a military coup. The observation period for the data we use in modeling starts in March 2001, and the last previous coup is thus outside our observation window. Restricting our knowledge to the observation period for our model data, we thus know that there has not been an  ILC in Thailand since March 2001, but we do not know how far back in time the last  ILC is. This issue is known as left-censoring, and ignoring it, e.g. by starting to count in March 2001, can lead to inaccurate model estimates. 

To ameliorate the issue, we go back further in time when looking for previous  ILCs, to January 1st, 1955. In other words, when building the variables necessary for duration modeling, we capture all  ILCs from 1955 on. Since the data are monthly, this drastically reduces the issue of left-censoring. For example, consider an extreme hypothetical example in which a country experienced  ILC in January 1955 and another one 555 months later in April 2001. By ignoring the issue we would code previous time to failure as 1 month, whereas our actual coding for this example would be 555, the true value.  

Obviously this does not entirely solve the issue, since it is still possible that a country experienced an  ILC before 1955. Twenty-five countries in our data have duration values equal to the maximum possible value, i.e. the $\sim$700 months covered by the period over which we build the duration data. Since it is unlikely that these countries experienced a  ILC in the last month of 1954, before our time window opens, they are probably left-censored and may have experienced an  ILC at some point in the past if we look back far enough.\footnote{Canada, Mexico, Costa Rica, United Kingdom, Ireland, Netherlands, Belgium, Luxembourg, France Switzerland, Germany Austria Italy Finland Norway Denmark Iceland Jordan Mongolia Taiwan North Korea, Japan, Bhutan, Australia, New Zealand} However, this is in practical terms not a large issue considering that the relationship between time to failure and covariates in our model is logarithmic. Thus while the natural logarithm of our maximum possible time to failure value is 6.56, adding another 50 years of observation would only increase the theoretical maximum to 7.18 and adding another 100 years beyond that, back to 1805, to 7.82. Appendix AB shows a comparison of two simple models with varying start dates for the duration data build to the same effect. 

To verify this, we estimated two version of a model using data built from 1955 on and 1840 on, respectively. The correlation between the duration variables in the the two data sets is 0.84, and the model estimates are consequently fairly similar. The estimates and model fit statistics are listed in detail below. The main point is that pushing the start date back does not dramatically change model fit nor regression point estimates, although it does seem to improve the precision for some variable coefficient estimates.\footnote{We also found during this process that the 2012 Archigos update we were provided with appears to be missing leaders for earlier time periods for some countries. For example, there are no recorded leaders for the Netherlands until 1875. This makes us hesitant to use 1840 as the start date for our data.}

Finally, we should also point out that left-censoring is not a problem unique to duration modeling. Most modeling efforts using country-year or country-month data, whether they use duration models or logistic regression or some other type of regression modeling, at least implicitly suffer from left-censoring as well. Existing work tends to either ignore duration dependence, which does not solve the problem, or if it uses some form of duration dependence control in the form of splines \citep{beck1998taking} or time polynomials, will suffer form the same left-censoring issue we face. 

In addition, work based on yearly observations arguably suffers more from this left-censoring issue than our monthly data do. For example, \citet{goldstone2010global} also use 1955 as the start date for their data, but with yearly observations there are by definition only one-twelfth the time-periods that monthly data would have. With yearly observations from 1955 to 2014 the maximum possible natural logarithm of time is 4.08, adding another 50 years increases it to 4.7, a ratio that is greater than the one with monthly data, and which by implication creates a greater potential of model bias due to left-censoring.

In summary, using 1955 as a start date for time to failure coding in our monthly data provides a time window large enough to render left-censoring a negligible issue.

\subsubsection{Short model, based on data with 1955 start date:}
\begin{verbatim}
Call:
spduration::spdur(
    duration = "duration ~ intratension.l1", 
    atrisk = "atrisk ~ log(NY.GDP.PCAP.KD.l1) + exclpop.l1 + 
        log(MS.MIL.XPND.GD.ZS.l1)", 
    data = train.short)

Duration equation: 
                 Estimate   StdErr      t      p    
(Dur. Intercept)  6.10721  0.37157 16.436 <2e-16 ***
intratension.l1  -0.01403  0.01066 -1.316  0.188    

Risk equation: 
                           Estimate    StdErr      t     p
(Risk Intercept)             1.6572 2742.2791  0.001 1.000
log(NY.GDP.PCAP.KD.l1)       1.9584  393.9210  0.005 0.996
exclpop.l1                   0.4240 1620.0521  0.000 1.000
log(MS.MIL.XPND.GD.ZS.l1)   -0.4991  243.8487 -0.002 0.998

           Estimate StdErr     t        p    
Weibull Shape $\alpha$   0.8996 0.1359 6.617 3.66e-11 ***
---
Signif. codes: *** = 0.001, ** = 0.01, * = 0.05, . = 0.1
\end{verbatim}

\subsubsection{Long model, based on data with 1840 start date:}

The direction of effects is the same as with the preceding short data model. The most notable difference is a dramatic increase in the certainty of estimates. 

\begin{verbatim}
Call:
spduration::spdur(
    duration = "duration ~ intratension.l1", 
    atrisk = "atrisk ~ log(NY.GDP.PCAP.KD.l1) + exclpop.l1 + 
        log(MS.MIL.XPND.GD.ZS.l1)", 
    data = train.long)

Duration equation: 
                 Estimate   StdErr      t      p    
(Dur. Intercept)  6.05103  0.36804 16.441 <2e-16 ***
intratension.l1  -0.01406  0.01061 -1.326  0.185    

Risk equation: 
                          Estimate  StdErr      t      p    
(Risk Intercept)            2.5809  0.2827  9.129 <2e-16 ***
log(NY.GDP.PCAP.KD.l1)      0.7880  1.8404  0.428 0.6685    
exclpop.l1                  0.8142  0.0619 13.154 <2e-16 ***
log(MS.MIL.XPND.GD.ZS.l1)  -1.2698  0.5786 -2.195 0.0282 *  

           Estimate StdErr     t       p    
Weibull Shape $\alpha$   0.8818 0.1339 6.583 4.6e-11 ***
---
Signif. codes: *** = 0.001, ** = 0.01, * = 0.05, . = 0.1
\end{verbatim}

\subsubsection{Comparison of model fit:}

The long and short data model versions have very similar fit performance; the longer performs slightly worse.

\begin{verbatim}
#     1955 short data

         Brier       AUC       Brier       AUC
Train 0.896669 0.6790222 0.001776146 0.7719595

#     1840 long data

          Brier      AUC       Brier       AUC
Train 0.8938463 0.655285 0.001776285 0.7711462
\end{verbatim}

\subsection{CAMEO codes and quad categories}

The \href{http://eventdata.parusanalytics.com/cameo.dir/CAMEO.Manual.1.1b3.pdf}{codebook for CAMEO}, available from Parus Analytics, lists the codes used by CAMEO to categorize events. There are 20 codes (root-codes) for general types of events, like \texttt{MAKE A PUBLIC STATEMENT} or \texttt{PROTEST}, and each in turn contains a number of further sub-codes describing more specific types of events within the larger category, e.g. \texttt{Decline comment} or \texttt{Make positive comment}. 

The 20 root codes in turn are further grouped into 4 aggregate groups, so called \emph{quad categories}, that are defined by the combination of two dimensions: verbal versus material events on one dimension, and cooperate versus conflictual events on the other dimension. Table \ref{tab:quads} shows the specific root codes associated with each quad category. 

\begin{table}[htdp]
\caption{CAMEO root codes and quad categories}
\label{tab:quads}
\begin{center}
\begin{tabular}{lrl}
Quad Category & \multicolumn{2}{l}{Root code} \\ \midrule \midrule
Verbal cooperation & 01 & Make a public statement \\
& 02 & Appeal \\
& 03 & Express intent to cooperate \\
& 04 & Consult \\
& 05 & Engage in diplomatic cooperation \\ \midrule
Material cooperation & 06 & Engage in material cooperation \\
& 07 & Provide aid \\
& 08 & Yield \\ \midrule
Verbal conflict & 09 & Investigate \\
& 10 & Demand \\
& 11 & Disapprove  \\
& 12 & Reject \\
& 13 & Threaten \\ \midrule
Material conflict & 14 & Protest \\
& 15 & Exhibit military posture \\
& 16 & Reduce relations \\
& 17 & Coerce \\
& 18 & Assault  \\
& 19 & Fight \\
& 20 & Engage in unconventional mass violence \\ \midrule
\multicolumn{3}{l}{Source: \href{http://eventdata.parusanalytics.com/cameo.dir/CAMEO.Manual.1.1b3.pdf}{CAMEO Codebook 1.1b3}}
\end{tabular}
\end{center}
\label{default}
\end{table}%

The quad variables we construct with this scheme are simple a count of the number of events matching the quad category root codes in a country in a given month. We filter actors to ensure that both actors and the event take place in the country itself, i.e. all events are domestic.

\subsection{Variable descriptions}

\begin{itemize}
\item  \texttt{irr.entry, irr.exit, irr.t}: Variables coding whether a leader entered or lost power in that country in that month. The last variable codes ILC, i.e. whether either or both an irregular entry and exit occurred in a country-month.
\item \texttt{i.[matl|verb].[conf|coop].[GOV|DIS|REL]t[GOV|DIS|REL].l1}:, \\ e.g. \texttt{i.matl.coop.GOVtGOV.l1} are behavioral variables generated by the Duke team from the ICEWS event data that capture the reported number of events between two actors and where the events belong to a particular quad category (see above). Thus\\ \texttt{i.matl.coop.GOVtGOV.l1} is a country-month count of reported material cooperation evens from government actors to other government actors, lagged by 1 month. Events are filtered to ensure that both actors are in the same country, i.e. domestic interactions only. These variables are directional, and thus \texttt{\ldots DIStGOV.l1} captures events from dissident actors to government actors while \texttt{\ldots GOVtDIS.l1} captures events from government actors towards dissident actors. Some aggregations are less specific, i.e. \texttt{i.protest.tGOV.l1} captures all protest events directed at government actors in a country in a given month.
\item \texttt{duration, atrisk, cured, failure, t.0, end.spell}: Variables related to the duration models. ``Duration'' is the time to failure or censoring, ``end.spell'' indicates the end of a spell, due either to censoring or failure and ``atrisk'' and ``cured'' are indicators for spells that do or do not end in failure, respectively.
\item Spatial weights of other variables are named with the following scheme: (1) prefix of \texttt{W}, followed by (2) the method by which spatial weights were constructed, i.e. \texttt{centdist.std} for inverse distance between state centroids, \texttt{knn4} for the 4 nearest neighboring states, \texttt{gower.events} for Gower event-based dissimilarity, \texttt{gower.pol} for Gower Polity-based dissimilarity, and \texttt{gower.econ} for economic dissimilarity, and finally (3) the variable from which the spatial weight is constructed, i.e. \texttt{protest.tGOV.l1} for protests in surrounding countries in the past month. 
\item \texttt{Amnesty.l1} Amnesty International's \href{http://www.politicalterrorscale.org/}{Political Terror Scale}, a 1 to 5 scale that measures the level of political violence and terror a country experiences, based on Amnesty International annual country reports and US State Department Reports on Human Rights practices. 
\end{itemize}

\section{Model fit and forecast accuracy}

Our goal is to predict the occurrence of ILC in particular country-months, where occurrence is measured as either 0 or 1. From a modeling perspective this task is equivalent to binary classification, in which we try to map 0's and 1's generated from a model to a stream of 0's and 1's observed in historic data. Table \ref{tab:class} shows how the possible combinations of observed and predicted values from a model are commonly termed. There are several terms summarizing the ratios of different classification errors, partly based on whether 0's or 1's are more important to predict correctly.

\begin{table}[ht]
\caption{Model classification errors} \label{tab:class}
\begin{tabular}{ rr|c|c| }
\multicolumn{2}{r}{} & \multicolumn{2}{c}{Model} \\
\multicolumn{2}{r}{}
 &  \multicolumn{1}{c}{0}
 & \multicolumn{1}{c}{1} \\
\cline{3-4}
\multirow{2}{*}{Observed} & 0 & True Negative & False Positive \\
\cline{3-4}
& 1 & False Negative & True Positive \\
\cline{3-4} 
\multicolumn{4}{c}{} \\
 \multicolumn{4}{l}{True positive rate (sensitivity, recall) $=$ TP / (FN + TP)} \\
 \multicolumn{4}{l}{True negative rate (specificity) $=$ TN / (TN + FP)} \\
 \multicolumn{4}{l}{False positive rate (1-specificity) $=$ FP / (TN + FP)} \\
 \multicolumn{4}{l}{Precision $=$ TP / (TP + FP)} 
\end{tabular}
\end{table}

\subsection{Accuracy and the accuracy paradox}

At first glance, assessing the performance of a binary classification model is simple. We can measure simple accuracy as the ratio of correct predictions ($0\rightarrow0, 1\rightarrow1$) to the total number of observations. This is the sum of true positives and true negatives divided by the overall number of cases. Incidentally, in order to do this, we need to choose a cut point with which to bin the continuous 0 to 1 model predictions into discrete 0 or 1 values. Throughout this paper, we choose cut points for each model so as to maximize the F-score, which is discussed in more detail below.

Table \ref{tab:fit.cm.and.cy} shows the accuracy, as well as other statistics, for the thematic and ensemble models. The top half shows it for the original country-month predictions, the bottom half for annualized, i.e. country-year, predictions. We include the latter to facilitate direct comparison with other work at the country-year level. The cut points used for each model are shown by $\tau$. Accuracy consistently exceeds 95\%. 

While the overall classification accuracy appears to be a simple and sufficient statistic for summarizing model fit, it becomes problematic if there is imbalance in the binary variable that is modeled. Our data, like all conflict data, are severely imbalanced with a very low rate of 1's at 17 hundredths of 1 percent. Even if we were to annualize our data, the positive rate would still be under 2 percent. 

This leads to a phenomenon known as accuracy paradox, where a model with no predictive power has a higher accuracy than a model with some predictive power. For example, consider the model predictions from our ensemble over the whole date range we cover using a probability cut point of 0.0043 in Table \ref{tab:predlow}. These predictions have an accuracy of 98.8\%. An alternative model based on predicting 0 for all cases, which has no predictive power whatsoever since it cannot predict any positive cases, has an accuracy of 99.8\% simply because the number of 0's in the data is so large.

\begin{table}[ht]
\centering
\begin{tabular}{rrr}
  \hline
 & 0 & 1 \\ 
  \hline
0 & 23505 & 2199 \\ 
  1 &  22 &  22 \\ 
   \hline
\end{tabular}
\caption{Ensemble predictions, 2001 to 2014, $\tau = 0.0043$} 
\label{tab:predlow}
\end{table}

There are several alternative measures and tools that are more suitable for imbalanced data. We will cover four of them: recall, precision, the ROC curve, and AUC.

\subsection{Recall and precision, revised precision estimate with fuzzy matching}

Recall and precision measure the rate of observed positives identified by a model and the rate of cases identified by a model that are in fact observed positives, respectively. They are calculated, respectively as:

\begin{eqnarray*}
Recall & = & TP / (TP + FN) \\
Precision & = & TP / (TP + FP)
\end{eqnarray*}

While recall can tell us how good a model is at identifying observed positive cases, precision tells us how much stock we can put in model predictions overall. For the classification table shown in Table \ref{tab:predlow}, the recall is 50\%, meaning the model identifies almost half of the ILCs between 2001 and 2014. However, this comes at a large cost in false positives, and the precision as a result is 0.99\%. 

Using a different, higher threshold, i.e. a more conservative model, we obtain the classifications shown in Table \ref{tab:predoptim}. The recall and precision rates here are 9.1\% and 4.9\% respectively.

\begin{table}[ht]
\centering
\begin{tabular}{rrr}
  \hline
 & 0 & 1 \\ 
  \hline
0 & 25627 &  77 \\ 
  1 &  40 &   4 \\ 
   \hline
\end{tabular}
\caption{Ensemble predictions, 2001 to 2014, cut point 0.016} 
\label{tab:predoptim}
\end{table}

The Recall and Prec. columns in Table \ref{tab:fit.cm.and.cy} show the recall and precision for the ensemble and thematic models, both in the country-month and annualized country-year data. Although the rates are low in the country-month data, they increase substantially when annualized. The fact that they are low in the country-month data is simply a reflection of the increasing difficulty of the classification problem due to the larger sparsity in the data. 

However, to further investigate the tradeoff between precision and accuracy in the monthly data we investigated how many of the false positives for a given cut point value correspond to false positives that occurred within 6 months of an observed ILC. In other words, this tells us how many of the model's false positives are within the 6-month time frame we use for forecasting. For the threshold reported in Table \ref{tab:predlow}, 75 of the false positives fall within 6 months of an ILC, giving a revised precision of 4.37\%. For Table \ref{tab:predoptim}, 38 of the false positives are within 6 months for a revised precision of 52\%. 

\subsection{Accuracy revisited, balanced F-score}

As should be apparent, accuracy is a problematic measure of classification in imbalanced (and in our case sparse) data. Precision and recall are more sensitive to imbalances. To obtain a single measure of fit, we can combine the precision and recall in a balanced accuracy measure that penalizes for data imbalances. One way to do so is with the F-score, which is calculated as:

\begin{eqnarray*}
F_\beta & = & (1+\beta^2) * \frac{precision*recall}{precision*\beta^2+recall}
\end{eqnarray*}

The term $\beta$ measures how much more weight one places on recall over precision. We use $\beta=1$ throughout, i.e. getting true positives is as important as avoiding false positives. The F-score is the basis for all thresholds we use in the tables reported in this paper. 

\subsection{Separation plots}

\begin{figure}
\caption{Separation plots}
\label{fig:sep}
        \centering
        \begin{subfigure}[b]{0.48\textwidth}
                \caption{Calibration}
                \includegraphics[width=\textwidth]{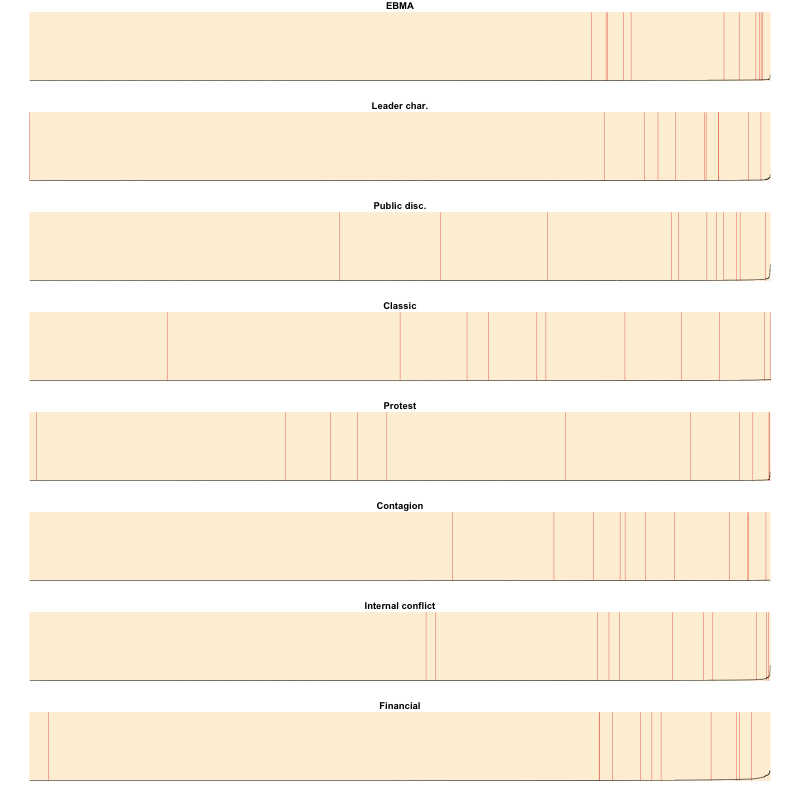}

        \end{subfigure}%
  	~
        \begin{subfigure}[b]{0.48\textwidth}
                \caption{Test}
                \includegraphics[width=\textwidth]{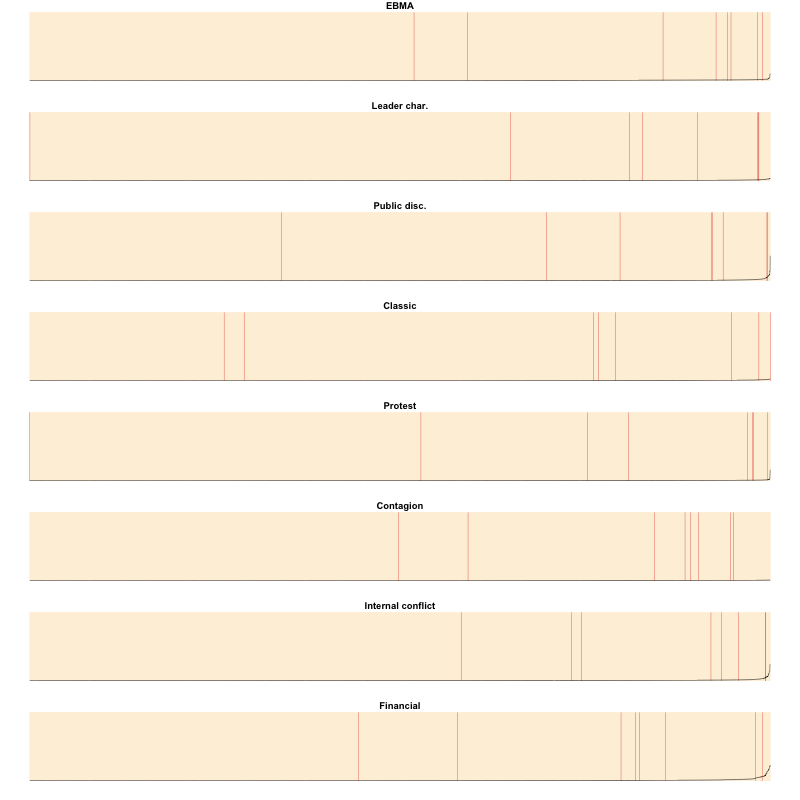}
        \end{subfigure}
\end{figure}

Separation plots \citep{sepplot:ajps:2011} provide a convenient visual method for assessing model fit. Figure \ref{fig:sep} shows the separation plots for the ensemble and thematic models in the calibration and test data. Each plot shows the country-months in the respective data as vertical bars that are colored tan for non-events and red for positive events, and orders them from left to right by their model predicted conditional hazard. The latter is also shown by the black line tracing the plot. Ideally, all red bars would be to the right of the plot, meaning they are assigned higher probabilities that the non-event cases in tan. 

\subsection{ROC and AUC}

One problem with the statistics discussed so far---accuracy, precision, recall, and the F-score---is that they depend on a particular choice of threshold value for binning the continuous model predictions into discrete 0 or 1 values that we can compare to the observed outcomes. As is common in models of rare events like irregular exit, the range of model probabilities is fairly low, with a high of 46 percent for a country-month. To convert these probabilities to 0 or 1 discrete values, logit and probit regression models usually use a threshold of 0.5. Given our low predicted values, we have to use a lower threshold, but the choice is still arbitrary. Using a threshold of 0.01, the ensemble predictions for the test period identify 3 country-months with an ILC at the cost of 39 false positives (country-months without an ILC). If we lower the threshold to 0.0025, the model identifies 9 of 11 ILCs, but at a cost of 929 false positives. As a result the accuracy, precision, recall, etc. static would change as well depending on which threshold we adopt. Thus, although the choice of a threshold for binning model predictions into 0 or 1 values is arbitrary, it greatly influences a model's classifications and therefore related summaries like accuracy, precision, recall, false positive rates, etc. 

Furthermore, there is a tradeoff between correctly identifying events and false alarms. A higher threshold results in a more conservative model with few false alarms, but it will also have fewer correct positives. Conversely, a lower threshold identifies more positives, but at the cost of more alarms. 

\begin{figure}[h!]
\caption{Forecast model ROC curve}
\centering
\includegraphics[width=4in]{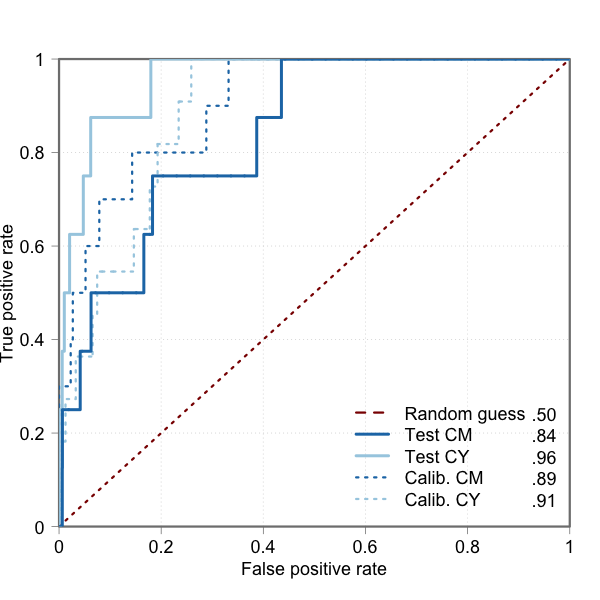}
\label{fig:roc}
\end{figure}

A ROC---receiver-operating characteristic---curve and the area under it (AUC) are a way to evaluate this tradeoff in a model between recall and precision, regardless of the threshold value. The ROC curve is a plot of the true positive rate (sensitivity, recall) against the false positive rate ($1 -$ specificity). Figure \ref{fig:roc} shows the ROC curves for the Ensemble model in the calibration and test sets, in both the country-month and annualized versions. The false positive rate is on the x-axis, and the true positive rate (recall) on the y-axis. Lines trace the performance of a model across different threshold values. For any model, we can increase true positives by increasing false positives. We are essentially buying true positives at a cost in false positives. 

In a model based on randomly guessing, this tradeoff is on average even, leading to the ROC curve shown in red. Thus, for example, if we randomly guess that 1 in every 67 cases is positive, which corresponds to a probability of roughly 0.015, we would on average get about 30 percent of ILCs right, but would also falsely classify 30 percent of the non-event country-months as positive. If we guessed 1 in every 12 cases to be ILCs, a probability of roughly 0.08 percent, the true and false positive rates would on average both be around 22 percent. Any useful model should thus exceed the random guess model. 

Although the tradeoff between precision and recall is present in all practical models, some can ``buy'' true positives at a lower cost in false positives. In a perfect model, in fact, all predictions would be true positives, with no false positives. The corresponding ROC curve would move from the bottom left corner to the top left and the top right. Models are better the closer they are to this ideal curve, i.e. the more true positives they can identify for a given cost in false positives. The ROC curves for the ensemble are well-performing. Unsurprisingly, they are better when annualizing the data, but even the country-month curves are much closer to the ideal point than the random guess model. 

The shape of the ROC curve is determined by the number and placement of positives in respect to the model probabilities. This can make it difficult to compare ROC curves across models. To summarize them, one can calculate the area under the ROC curve, i.e. the AUC. An ideal model would have a 100 percent true positive rate, and a 0 percent false positive rate. In the figure, it would just be a point in the top left corner, with an AUC of 1. The random guess, red line, has an AUC of 0.5. The AUCs for our models, 0.86 to 0.96, thus tell us that the model performs well. 

Table \ref{tab:fit.cm.and.cy} shows the AUC values for the ensemble and thematic models in all data partitions and for both the country-month and annualized data. 


\begin{sidewaystable}[ht]
\centering \footnotesize
\begin{tabular}{lr|rrrrr|rrrrr|rrrrr}
  \midrule
  \multicolumn{17}{l}{Country-month (original) predictions} \\
\midrule
  && \multicolumn{5}{c}{Train} & \multicolumn{5}{c}{Calibration} & \multicolumn{5}{c}{Test} \\ 
  \midrule
Model & W & AUC & $\tau$ & Acc. & Recall & Prec. & AUC & $\tau$ & Acc. & Recall & Prec. & AUC & $\tau$ & Acc. & Recall & Prec. \\ 
  \midrule
Ensemble &  & 0.845 & 0.017 & 0.997 & 0.080 & 0.071 & 0.890 & 0.013 & 0.993 & 0.182 & 0.077 & 0.839 & 0.016 & 0.992 & 0.250 & 0.074 \\ 
  Leader char. & 0.006 & 0.803 & 0.010 & 0.987 & 0.200 & 0.025 & 0.816 & 0.010 & 0.986 & 0.091 & 0.019 & 0.776 & 0.014 & 0.983 & 0.375 & 0.047 \\ 
  Public disc. & 0.007 & 0.848 & 0.024 & 0.995 & 0.120 & 0.049 & 0.832 & 0.026 & 0.992 & 0.091 & 0.037 & 0.820 & 0.046 & 0.994 & 0.250 & 0.111 \\ 
  Global Inst. & 0.006 & 0.779 & 0.007 & 0.964 & 0.200 & 0.008 & 0.717 & 0.023 & 0.998 & 0.091 & 1.000 & 0.726 & 0.025 & 0.998 & 0.125 & 1.000 \\ 
  Protest & 0.006 & 0.746 & 0.009 & 0.985 & 0.200 & 0.020 & 0.658 & 0.019 & 0.997 & 0.182 & 0.222 & 0.752 & 0.015 & 0.995 & 0.125 & 0.071 \\ 
  Contagion & 0.681 & 0.824 & 0.022 & 0.998 & 0.040 & 0.200 & 0.839 & 0.006 & 0.992 & 0.091 & 0.036 & 0.814 & 0.004 & 0.885 & 0.625 & 0.011 \\ 
  Internal conflict & 0.290 & 0.788 & 0.031 & 0.997 & 0.080 & 0.062 & 0.829 & 0.038 & 0.993 & 0.182 & 0.087 & 0.743 & 0.045 & 0.992 & 0.250 & 0.077 \\ 
  Financial & 0.006 & 0.782 & 0.074 & 0.999 & 0.040 & 1.000 & 0.789 & 0.017 & 0.953 & 0.273 & 0.014 & 0.787 & 0.037 & 0.979 & 0.250 & 0.026 \\ 
  \midrule
  \multicolumn{17}{l}{Country-year (annualized) predictions} \\
  \midrule
  && \multicolumn{5}{c}{Train} & \multicolumn{5}{c}{Calibration} & \multicolumn{5}{c}{Test} \\ 
  \midrule
Model & W & AUC & $\tau$ & Acc. & Recall & Prec. & AUC & $\tau$ & Acc. & Recall & Prec. & AUC & $\tau$ & Acc. & Recall & Prec. \\ 
  \midrule
Ensemble &  & 0.944 & 0.085 & 0.978 & 0.480 & 0.375 & 0.906 & 0.132 & 0.986 & 0.300 & 1.000 & 0.959 & 0.076 & 0.982 & 0.500 & 0.444 \\ 
  Leader char. & 0.006 & 0.835 & 0.087 & 0.981 & 0.520 & 0.448 & 0.775 & 0.084 & 0.982 & 0.400 & 0.571 & 0.750 & 0.111 & 0.982 & 0.500 & 0.444 \\ 
  Public disc. & 0.007 & 0.927 & 0.153 & 0.982 & 0.240 & 0.462 & 0.845 & 0.127 & 0.978 & 0.400 & 0.444 & 0.929 & 0.236 & 0.986 & 0.250 & 0.667 \\ 
  Global Inst. & 0.006 & 0.920 & 0.045 & 0.910 & 0.760 & 0.130 & 0.902 & 0.112 & 0.978 & 0.200 & 0.400 & 0.919 & 0.138 & 0.982 & 0.250 & 0.400 \\ 
  Protest & 0.006 & 0.924 & 0.086 & 0.981 & 0.640 & 0.457 & 0.923 & 0.059 & 0.986 & 0.600 & 0.667 & 0.986 & 0.069 & 0.994 & 0.875 & 0.778 \\ 
  Contagion & 0.681 & 0.930 & 0.104 & 0.980 & 0.240 & 0.375 & 0.848 & 0.036 & 0.949 & 0.400 & 0.174 & 0.911 & 0.057 & 0.974 & 0.250 & 0.222 \\ 
  Internal conflict & 0.290 & 0.937 & 0.112 & 0.982 & 0.400 & 0.476 & 0.891 & 0.073 & 0.970 & 0.800 & 0.381 & 0.950 & 0.292 & 0.990 & 0.375 & 1.000 \\ 
  Financial & 0.006 & 0.881 & 0.185 & 0.982 & 0.200 & 0.417 & 0.896 & 0.270 & 0.974 & 0.300 & 0.333 & 0.876 & 0.646 & 0.982 & 0.125 & 0.333 \\
   \midrule
   \multicolumn{17}{p{8in}}{$\tau =$ cut point value for binning to 0/1 values that maximizes the F-score for that model (harmonic mean of recall and precision).}
\end{tabular}
\caption{Ensemble and thematic model fit in country-month and annualized data} 
\label{tab:fit.cm.and.cy}
\end{sidewaystable}


\end{document}